\documentclass{emulateapj}

\def\ga{\mathrel{\hbox{\rlap{\hbox{\lower4pt\hbox{$\sim$}}}\hbox{$>$}}}}
\def\la{\mathrel{\hbox{\rlap{\hbox{\lower4pt\hbox{$\sim$}}}\hbox{$<$}}}}

\usepackage{wrapfig}
\usepackage{verbatim}
\usepackage{amsmath}
\usepackage{multirow}
\usepackage{relsize}
\usepackage{amssymb}
\usepackage{float}

\newcommand{\sigs}{\sigma_s}

\newcommand{\meanrho}{\rho_0}

\newcommand{\means}{s_0}

\newcommand{\cs}{c_\mathrm{s}}

\newcommand{\turbustat}{{\sc TurbuStat}}

\shorttitle{Diagnostics of The Turbulent ISM}
\shortauthors{Burkhart}

\begin{document}

\title{Diagnosing Turbulence in the Neutral and Molecular Interstellar Medium of Galaxies}

\author{Blakesley Burkhart}
\affiliation{Department of Physics and Astronomy, Rutgers, The State University of New Jersey, 136 Frelinghuysen Rd, Piscataway, NJ 08854, USA}
\affiliation{Center for Computational Astrophysics, Flatiron Institute, 162 Fifth Avenue, New York, NY 10010, USA}

\begin{abstract}
Magnetohydrodynamic (MHD) turbulence is a crucial component of the current
paradigms of star formation, dynamo theory, particle transport, magnetic reconnection
and evolution of structure in the interstellar medium (ISM) of galaxies. Despite the importance of turbulence to astrophysical fluids, a full theoretical framework based on solutions to the Navier-Stokes equations remains intractable.  Observations provide only limited line-of-sight information on densities, temperatures, velocities and magnetic field strengths and therefore directly measuring turbulence in the ISM is challenging. A statistical approach has been of great utility in allowing comparisons of observations, simulations and analytic predictions.
In this review article we address the growing importance of MHD turbulence in many fields of astrophysics and review  statistical diagnostics for studying interstellar and interplanetary turbulence. In particular, we will review statistical diagnostics and machine learning algorithms that have been developed for observational data sets in order to obtain information about the turbulence cascade, fluid compressibility (sonic Mach number), and magnetization of fluid (Alfv\'enic Mach number). These techniques have often been tested on numerical simulations of MHD turbulence, which may include the creation of synthetic observations, and are often formulated on theoretical expectations for compressible magnetized turbulence. We stress the use of
multiple techniques, as this can provide a more accurate indication of the turbulence parameters of interest. 
We conclude by describing several open-source tools for the astrophysical community to use when dealing with turbulence.

\end{abstract}

\section{Introduction: The Turbulent Universe}

Turbulence, a physical phenomenon in which a flow devolves into a cascade of swirls and eddies, is a ubiquitous state of both terrestrial and astrophysical fluids \citep{Kolmogorov41a,GS95,ElmegreenScalo,Lazarian07rev,krumreview2014}. More quantitatively, a ``turbulent flow'' can be distinguished from a smooth or ``laminar flow'' by a high Reynolds number. The Reynolds number is the ratio of inertia to
viscous forces in a fluid. 
and is named after the famous British fluid dynamicist Osborne Reynolds (1842 -1912). It is a dimensionless quantity $R_e=\frac{V{_L}L}{\nu}$, where V$_L$ is a characteristic velocity, L is a characteristic scale and $\nu$ is the fluid viscosity. 
Formally, the dynamics of a hydrodynamic turbulent flow with velocity $\textbf{u}$, density $\rho$, viscosity $\nu$ and acceleration $\textbf{F}$ should be described by the Navier-Stokes equations:
\begin{equation}
{\partial{\bf u}\over{\partial t}} + ({\bf u} \cdot \nabla) {\bf u} = - {1\over\rho} \nabla p + \nu\nabla^2{\bf u} + {1\over\rho}{\bf F} 
\end{equation}
with mass conservation:
$\partial{\rho}/\partial{t}+\nabla\cdot({\rho {\bf u}})=0$

Although the Navier-Stokes equations were written down in the 19th century and describe Newtonian motion for fluids, they remain largely analytically intractable for nonlinear fluid motions with high Reynolds numbers.  Standard partial differential equation (PDE) methods appear inadequate to settle the problem. Furthermore, turbulent fluids have a large number of degrees of freedom and exhibit chaotic behavior.  It is due to these difficulties that Richard Feynman notably called turbulence ``the most important unsolved problem in classical physics."  Indeed, our understanding of fluid turbulence is so inadequate and yet so important  that substantial progress toward a mathematical theory of turbulence using the 3D Navier-Stokes equations will garner a Clay Foundation Millennium Problem Prize\footnote{http://www.claymath.org/millennium-problems}.

The ``turbulence problem" in physics extends beyond the Earth to astrophysics. For astrophysical fluids in particular, the medium is often fully or partially ionized to the point where it becomes conducting and the dynamics introduced by a turbulent magnetic field become important.  In this case, magnetohydrodynamic (MHD) turbulence becomes one of the major processes that
govern the structure formation and evolution of the interstellar medium of galaxies \citep{Biskampbook,2019tuma.book.....B}. 
Partially and fully ionized fluid plasmas in space have very high Reynolds numbers \citep{ElmegreenScalo,Scalo04a,2013LRSP...10....2B,Xu2016b}. For example, consider cold molecular gas in the interstellar medium of galaxies: the kinematic viscosity is of order $\nu \sim 10^{16}\rm{cm}^2s^{-1}$, assuming a sound speed of $0.2 \rm{kms}^{-1}$ and a particle mean free path of $\lambda=10^{12} \rm{cm}$. For a typical galactic molecular cloud of size L=30pc and V$_L=3\rm{kms}^{-1}$, this gives $R_e \sim 10^9$, whereas the transition to a turbulent flow takes place around $R_e \sim 10^3$ \citep{doi:10.1098/rsta.1895.0004,doi:10.1098/rsta.2008.0217,book1,Mullin2011}. In the presence of magnetic fields, the ability of ions to move perpendicular to the magnetic field is inhibited and this further increases $R_e$ \citep{Braginskii:1965,YL12book}.

\begin{figure}
\includegraphics[width=10cm]{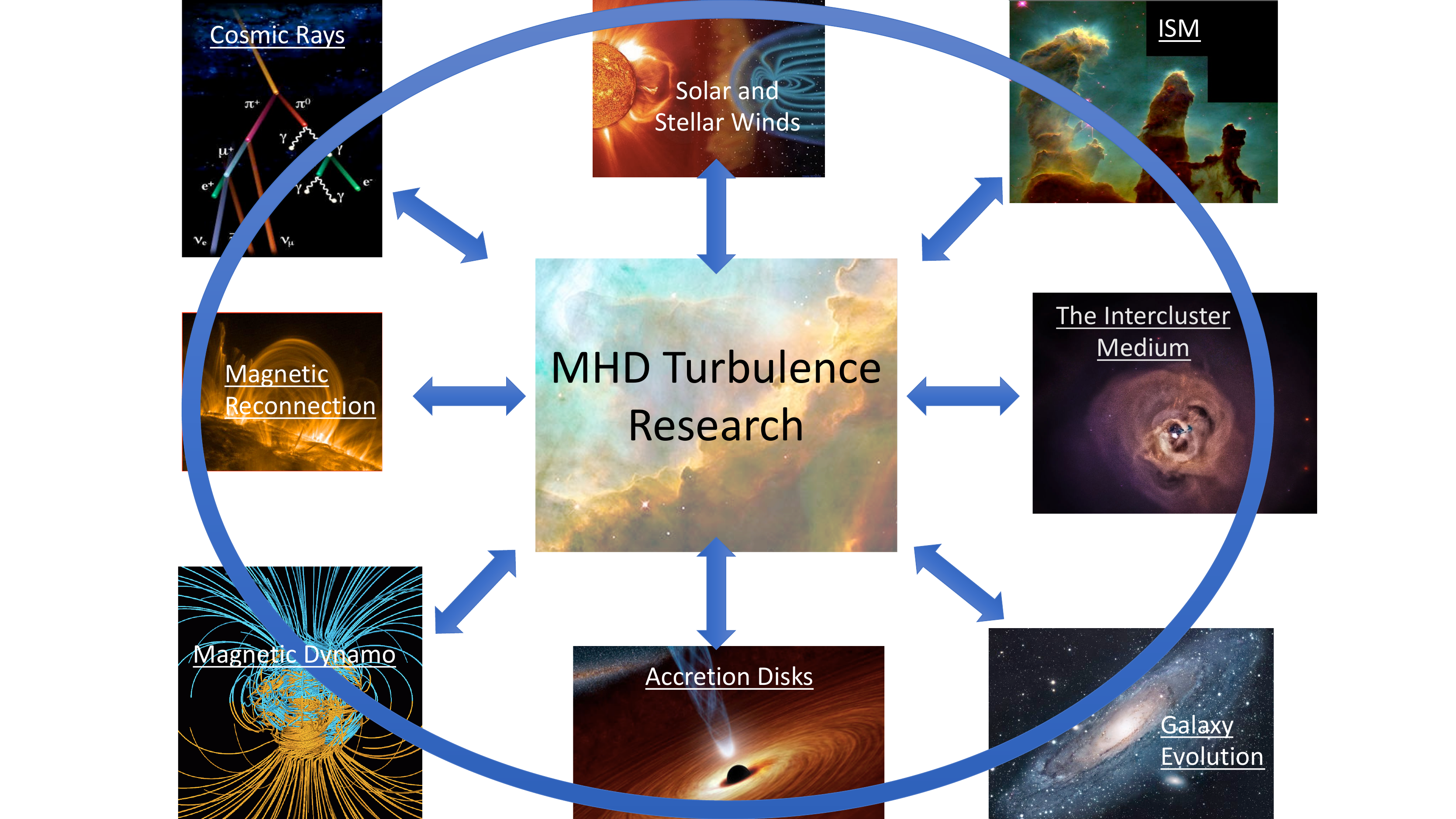}
\caption{
\label{fig:mhdcartoon}
A cartoon illustrating how many basic astrophysical processes and environments are linked intricately to our understanding of MHD turbulence. This cartoon does not contain a complete list. Progress in these individual areas of study hinges on the characterization of MHD turbulence properties. 
}
\end{figure}

MHD turbulence is therefore recognized as an important component of galaxy evolution and part of the established paradigm of the interstellar (ISM), interplanetary (IPM) and intercluster media (ICM)\citep{Armstrong95,ElmegreenScalo,San11,Schilick12,XuZ16}.
Turbulent eddies create density and magnetic field fluctuations which, in turn, can control the processes of star formation \citep{Lars81,Mckee_Ostriker2007,maclow04,ElmegreenScalo,Collins12a,padoan2017ApJ...840...48P,Gallegos-Garcia:2020:ApJL:,2021arXiv210202868P}, cosmic ray diffusion and acceleration \citep{Schlickeiser02,LY14,Xu2016b}, and the formation of accretion disks around planets/stars \citep{Balbus91a,Hughes10a,Ross2017} 
and black holes \citep{Johnson2016}.  It is now also recognized that turbulence in ionized plasmas controls the evolution of magnetic fields in the universe through the fields' creation via the turbulent dynamo and their evolution through reconnection diffusion \citep{1993IAUS..157..487K,LV99,2002PhRvL..89z5007B,Brandenburg2005,2008A&A...486L..35G}.  Turbulence also spatially redistributes magnetic fields via a process known as reconnection diffusion \citep{Santos-Lima12a}.  Magnetic fields in turn can inhibit the formation of stars \citep{Mestel65a,Mouschovias87a}   and control heat transport in galaxy clusters \citep{Narayan_Medv,Lazarian2006ApJ...645L..25L}.  
Turbulence in the ISM can be driven by a wide variety of energetic events acting on different scales.  On large (kiloparsec) scales, these drivers can include supernova explosions, magnetorotational instability, shear motions, and gravitational disk instabilities \citep{maclow04,kim06,10.1111/j.1365-2966.2008.14043.x,falceta14,2009AJ....137.4424T,Krumholz10c,Forbes14a}.  On intermediate and small scales (tens of parsec and below), these drivers include stellar winds and jets from new stars \citep{Offner09b,Cunningham12a}.
Figure \ref{fig:mhdcartoon} illustrates a few  of the galactic and extragalactic environments and processes in which turbulent gas motions are found to be important. 

\begin{figure*}
\includegraphics[width=15.5cm]{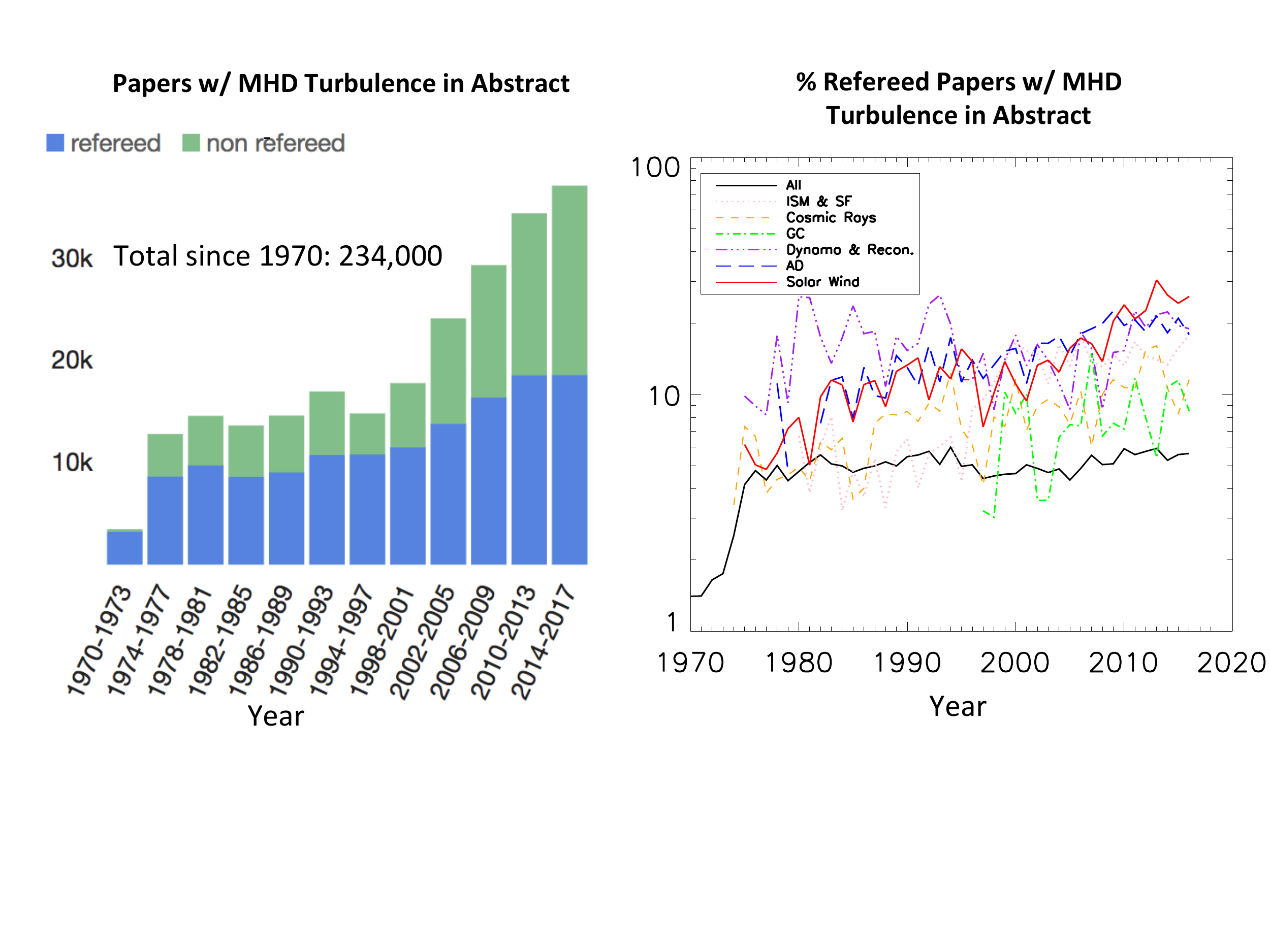}
\caption{
\label{fig:citations}
Left panel:  total number of publications that mention the words MHD or turbulence in the title or abstract, binned in three-year intervals.  Right panel: percent of refereed astronomy papers mentioning MHD turbulence in the abstract.
Total number of papers is shown with a black line. 
Subfields:  GC denotes the keywords ``Galactic Center"; ISM \& SF denotes the keywords ``Interstellar Medium and Star Formation"; AD denotes the keywords ``Accretion Disk"; Dynamo \& Recon. denotes the keywords ``dynamo" and ``reconnection".  
Source: ADS Bumblebee.}
\end{figure*}

 The importance of turbulence for astrophysics is further evident in the current and historical literature trends. The left panel of Figure \ref{fig:citations} shows the total number of publications that mention the words MHD or turbulence in the title or abstract, binned in three-year intervals from 1970-2017. Since 1970, more than  220,000 papers in the field of astronomy have been published with turbulence as a major component, focus, or conclusion of a study and this trend is still growing. The right panel of Figure \ref{fig:citations} shows the percent of refereed astronomy papers mentioning MHD turbulence in the abstract, broken down by keyword search. 
Astronomy subfields such as the study of ISM, accretion disks (including black holes and protoplanetary disks), and solar wind have seen a steady increase in the number of publications closely dealing with turbulence.

Despite the widely recognized importance of MHD turbulence for galactic processes, \textit{the astrophysical community faces a dilemma}. If MHD turbulence is such an important component of cutting-edge research in star formation, galaxy evolution, cosmic ray physics, etc., and yet no complete theoretical description of turbulence exists, can these various fields expect to make substantial progress? Furthermore, MHD turbulence is notoriously difficult to study in a controlled laboratory setting \citep{Nornberg2006}.  The situation becomes even more bleak when trying to quantify astrophysical turbulence when dealing with line-of-sight (LOS) convolution, radiative transfer effects, and telescope beam smearing \citep{Burkhart2013b}.
The solar wind community has the advantage of in-situ measurements to characterize turbulent density, velocity and magnetic field fluctuations and yet has still faced difficulty in devising metrics that can accurately characterize plasma fluid properties.

Fortunately, there are several different ways to make tractable progress in our understanding of turbulence in the interstellar medium of galaxies.  The purpose of this review paper is to provide an overview of key diagnostics for important parameters of turbulence as it is found in the dense and molecular ISM of galaxies.  These metrics have been inspired by a combination of numerical simulations and analytic theories of turbulence.   
Numerical simulations of MHD turbulence have been tremendously influential in our understanding of the physical conditions and statistical properties of MHD turbulence both in astrophysical environments  \citep{2002CoPhC.147..471B,maclow04,Ballesteros-Paredes07a,Mckee_Ostriker2007,PhysRevLett.114.075001,ZIEGLER2008227,Bendre_2020}  and in laboratory experiments \citep{Nornberg2006,Bayliss2007}.  In fact, the surge both in publications related to turbulence (see Figure \ref{fig:citations}) and in our understanding of turbulence as a key component of fluid astrophysics can, in part, be attributed to advances in computational astrophysics.
Present 3D numerical MHD turbulence codes can produce simulations that resemble observations in terms of structure (e.g., filaments and fractals in the ISM and star-forming regions),
star formation rates in clouds \citep{maclow04,Krumholz2005,Padoan11b,Hennebelle11a,federrath12,Collins12a,Lee2015ApJ...800...49L,Mocz2017},  cosmic ray physics \citep{Schlickeiser02,YL12book}, and galaxy gas velocity dispersions \citep[e.g.,][]{kim02,kim06,falceta14,bournaud14a}, to name just a few.
Adaptive mesh refinement (AMR) simulations have allowed increased resolution in studies of collapsing regions in a gravoturbulent flow. 
In order to resolve turbulence driven both by large scale processes as well as on the smaller scales by the gravitational potential during collapse, AMR is required. The energy injection scale of gravity-driven turbulence is close to the local Jeans scale and should be resolved to a minimum of 30 cells \citep{Federrath_2011}.
AMR methods have opened up a wide range of studies that investigate how turbulence affects star formation and galaxy evolution \citep[e.g.,][]{Collins12a,Rosen14a,Federrath2015,Mocz2017,padoan2017ApJ...840...48P,Grudic2018,orr2019MNRAS.486.4724O}.

At the same time, one can approach the turbulence problem from the point of view of analytic scalings. 
A statistical view of turbulence allows researchers to describe overall properties of the fluid flow. While turbulence appears chaotic to the eye, statistical averaged properties lend themselves to analytic reproducibility and regularity when averaged over spatial or temporal domains \citep{Kowal2007,Kritsuk07a}.  We will review basic analytic scalings of MHD turbulence in Section 2. When combined with statistical diagnostics, a comparison of numerical simulations with analytic prediction can determine the extent to which the numerical ``reality" represents actual physical processes.  Statistical diagnostics for turbulence properties can be designed for observations and tested on numerical simulations with a wide range of parameters that are anticipated for the ISM. 
In this review, we focus on the progress that has been made in the field of galactic astrophysical MHD turbulence from the point of view of statistical diagnostics applied to observational data, which can inform researchers on the properties of the flow. 

\begin{figure*}
\includegraphics[width=15.5cm]{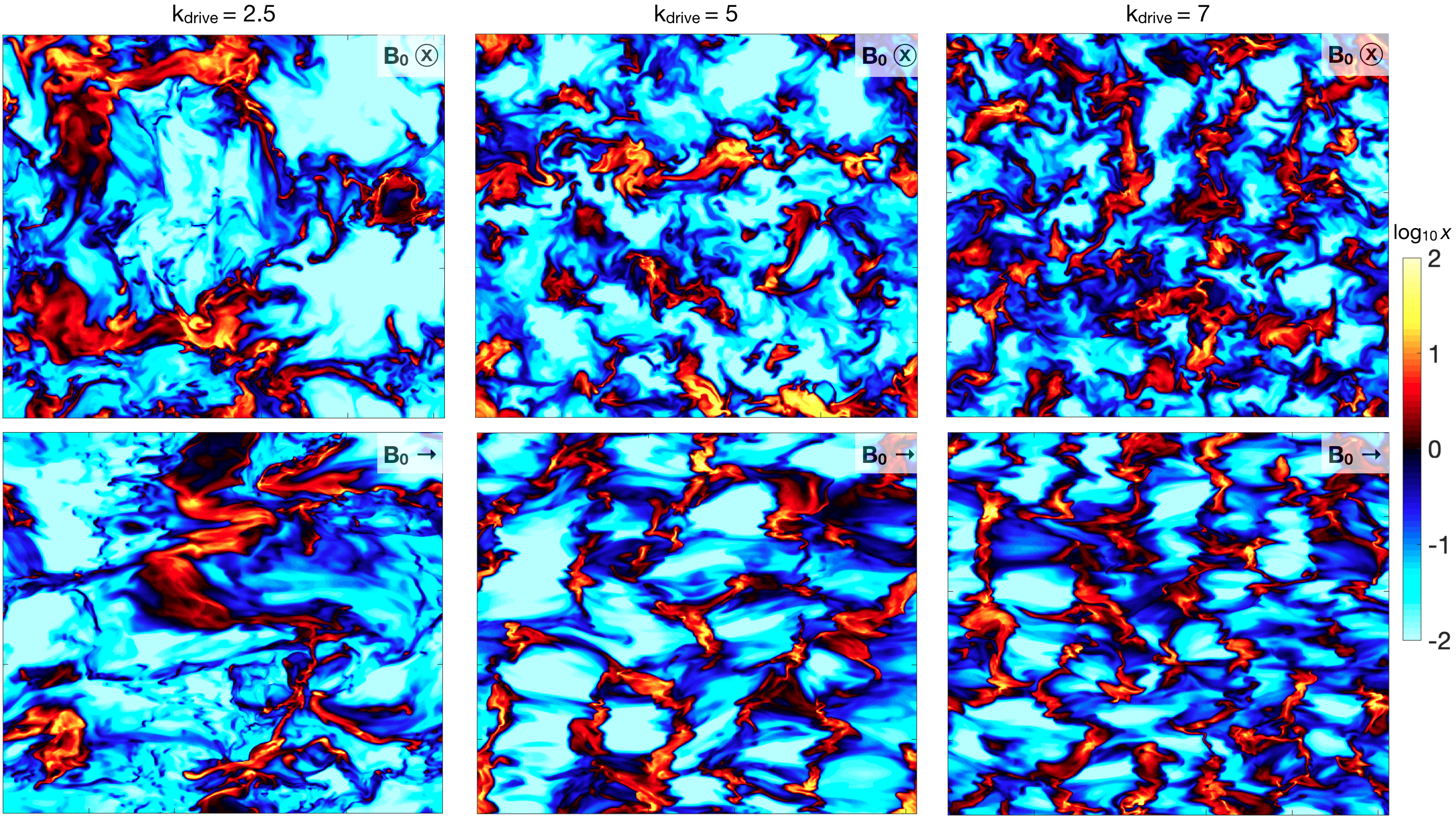}
\caption{
\label{fig:sims}
Example of MHD simulations reproduced from \citet{Bialy2020}: Density slices through MHD turbulence simulations with different driving scales. The driving scales are at $k=2.5$ (left), $k = 5$ (middle) and $k= 7$ (right). The top panel shows the mean magnetic field along the axis of density integration and the bottom panel shows the density integration perpendicular to the mean magnetic field. }
\end{figure*}

The review is organized as follows: 
In Section \ref{sec:theory}, we summarize progress that has been made in the study of MHD turbulence from the point of view of analytic predictions   and numerical simulations. We also give a brief overview of the incredible utility of numerical simulations in MHD turbulence research in the last decades and how it is of great importance to transform such simulations into synthetic observations.  In Section \ref{sec:stats}, we discuss how turbulence studies, whether observational or numerical, must be approached from the point of view of statistical diagnostics and we review key diagnostics that have recently been developed and applied to observations.  A full diversity of diagnostics 

Finally, in Section \ref{sec:pathforward}, we conclude by highlighting the best path forward for turbulence-related research in astrophysics: comparing the results of statistical techniques of turbulence, simulations and analytic predictions with observations. To this end, we point out a number of shared community resources and code repositories, which provide a way forward for different groups to approach the problem  of astrophysical turbulence from a variety of  perspectives.

\section{Theoretical Description of MHD Turbulence}
\label{sec:theory}

What are the important and basic aspects of a turbulent flow in the ISM which we would like to measure? In order to address this question it is important to review the properties of turbulence and the progress in understanding its analytic scaling.
In this section, we will provide an overview of theoretical progress in MHD turbulence.  Analytic theory both informs diagnostic tools as well as provides a point of reference to compare observations and numerical simulations to.  We stress that it is not the purpose of this paper to review MHD turbulence theory. There are many excellent reviews and books devoted to the theory of MHD turbulence \citep[e.g.][]{Biskampbook}.  Rather, the purpose here is to motivate the theoretical basis for the statistical techniques that will be introduced later on.

\subsection{Analytic Theories of Incompressible MHD Turbulence}

As a precursor to discussing MHD turbulence it is important to review the basics of hydrodyanmic turbulence.
The drivers of turbulence
 inject energy at large scales and this  energy cascades down to smaller and smaller scales via a hierarchy of eddies. A key hydrodynamic theory of turbulence is the famous Kolmogorov (1941) theory. Here energy is injected at large scales, creating large eddies corresponding to large $R_e$. Because $R_e>>1$, the energy transfer continues to progressively smaller eddies until finally $R_e\approx1$.  At this point, the smallest eddies dissipate their energy on the timescale of a local eddy turnover time.

The incompressible hydrodynamic cascade of energy goes as $\sim v^2_l/\tau_{casc, l} =const$, where $v_l$ is the velocity at scale $l$. The turnover time for the eddies of size $l$ is $\tau_{cask, l}\approx l/v_l$. From this, the well-known relation $v_l\sim l^{1/3}$ follows.
This corresponds to a Fourier power spectrum  scaling  as $E(k) \propto k^{-5/3}$,  where $k$ is  the wavenumber.
This predicted scaling is important for testing the numerical accuracy of simulations and for comparing velocity statistics of turbulence, as will be discussed in the next section.

Interestingly, in the case of incompressible MHD turbulence, the power spectrum does not deviate strongly from the \citet{Kolmogorov41a} scaling.  However, in contrast with the isotropic nature of Kolmogorov turbulence, in the presence of a dynamically important magnetic field, eddies become anisotropic. A number of theoretical arguments have been made for this; perhaps most famous is the \citet{GS95} theory, henceforth referred to as GS95.
GS95 considers turbulent eddies in relation to their orientation perpendicular or parallel to the mean magnetic field.
An important aspect of the MHD cascade is that scales $l_{\bot}$ and $l_{\|}$ should be measured with respect to the direction of the local magnetic field \citep{LV99}. 
An ionized magnetized fluid element feels the ``tug" of the magnetic field, which induces \textit{scale-dependent anisotropy} along the direction of the magnetic field.  The turbulent energy cascade proceeds in the perpendicular direction to the mean field and the velocity perpendicular to magnetic field direction scales with length $l_{\perp}$ as $v_{l} \sim l_{\perp}^{1/3}$.  The mixing motions perpendicular to the field induce Alfv\'enic perturbations that determine the parallel size of the magnetized eddy.  In the GS95 scenario this is known as
{\it critical balance}, i.e., the equality of the eddy turnover time $l_{\bot}/v_l$ and the period of the corresponding Alfv\'en wave $\sim l_{\|}/V_A$, where $l_{\|}$ is the parallel eddy scale and $V_A$ is the Alfv\'en velocity. Making use of the earlier expression for $v_l\sim \l_{\bot}^{1/3}$, one  obtains $l_{\|}\sim l_{\bot}^{2/3}$, which demonstrates that eddies become increasingly elongated as the energy cascades proceed to smaller scales \citep{Cho2002,Bere15}.
\textit{The scale-dependent anisotropy is the keystone for statistical techniques that measure the strength and orientation of the magnetic field, as will be discussed later.}

GS95 theory postulates that the injection of energy at scale $L$ is isotropic and the injection velocity is equal to the Alfv\'en velocity  $V_A$ in
the fluid; i.e., the Alfv\'en Mach number $M_A\equiv (V_L/V_A)=1$, where $V_L$ is the injection velocity. Thus GS95 theory provides the description
of trans-Alfv\'enic turbulence. This model was later generalized
for both the sub-Alfv\'enic case, i.e., $M_A<1$, and the super-Alfv\'enic case, i.e., $M_A>1$ \citep{LV99}.  Indeed, if $M_A>1$, instead of using the driving scale, $L$,  one can use another scale, $l_A$:
\begin{equation}
l_A=L(V_A/V_L)^3=LM_A^{-3}
\label{eq:alf1}
\end{equation}
which is the scale at
which the turbulent velocity equals $V_A$.  When $M_A\gg 1$,
 turbulence is expected to follow the isotropic
 Kolmogorov cascade $v_l\sim l^{1/3}$ over the
range of scales $[L, l_A]$. 
When $M_A<1$, turbulence obeys GS95 scaling (this is also called ``strong''
MHD turbulence) not from the largest scale $L$ but from a smaller scale ($l_{trans}$): 
\begin{equation}
l_{trans}\sim L(V_L/V_A)^2\equiv LM_A^2
\label{eq:trans1}
\end{equation}
 While in the range $[L, l_{trans}]$, the turbulence is ``weak.''\footnote{The notion ``strong'' should  be associated with the strength of the nonlinear interactions rather than the amplitude of the turbulence. } 

The purpose of the above discussion is to highlight how MHD turbulence connects the statistics of velocity and magnetic fluctuations. For this review on observational statistics of turbulence, this connection becomes incredibly important as it makes it possible to diagnose the magnetic state of a fluid via velocity and density statistics, as we will later show. 
However, we would be remiss not to alert the reader to additional and modified scaling relations developed in the time since GS95.  For example, the theory of dynamic alignment  predicts a -3/2 power-law slope scaling as opposed to the GS95 scaling of -5/3  for incompressible MHD turbulence \citep{Boldyrev,Boley06a,Perez13a}.  The difference in these scalings has been controversial \citep{BeL10,2013arXiv1301.7425B,2014ApJ...784L..20B}, even when measured using the highest-resolution numerical simulations available \citep{2021NatAs.tmp...13F}, and there is currently no possibility of discerning these slopes from ISM observations. As this review focuses on statistics of turbulence, particularly on observations of \textit{compressible} MHD turbulence, we  will  not focus on the differences between these models further.

\subsection{Analytic Theories of Compressible Turbulence and the Effect of Density Fluctuations}
The previous overview of MHD turbulence theory only considered incompressible Alfv\'enic turbulence.  Alfv\'enic turbulence develops an independent cascade which is marginally affected by the fluid compressibility. Compressible modes can develop their own cascade   \citep[ e.g., fast mode cascade, ][]{cho03}. Astrophysical fluids are often highly compressible, with high sonic Mach numbers \footnote{The ratio of the turbulent velocity to the sound speed, known as the sonic Mach number, characterizes the compressibility of the medium.}, and compressibility introduces a number of very important effects.  These include strong density fluctuations from shocks, which produce lognormal density distributions and shallower than \citet{Kolmogorov41a} power spectrum slopes for density fields \citep{Beresnyak05a}.

Attempts to include effects of compressibility in analytic 
 descriptions of turbulence can be dated as far
back as the 1950s with the the work of \citet{1951ApJ...114..165V}, which used a model based on hierarchical clustering of gas where large clouds consisted of increasingly smaller clouds as one moves to smaller scales. 
 \cite{1951ApJ...114..165V} then derived a relationship between subsequent levels of hierarchy
\begin{equation}
 {\rho _\nu}/{\rho_{\nu - 1}} = \left( {l_\nu}/{l_{\nu - 1}} \right )^{-3\alpha} ,
 \label{eqn:von}
\end{equation}
where $\rho_\nu$ is the average density inside a cloud at level $\nu$, $l_\nu$ is the mean size of the cloud and $\alpha$ is constant that quantifies compression at each level.

Later \citet{Fleck1996} extended this hierarchical clump model with energy transfer in compressible fluid to obtain the scaling relations for compressible turbulence by combining \ref{eqn:von} with the understanding that, in a compressible fluid, the volume energy transfer rate is constant in a statistical steady state, e.g., 
 $\varepsilon_V = \rho \varepsilon \sim \rho v^2/(l/v) = \rho v^3/l.$
 The hierarchical fluctuations of velocities in the \citet{Fleck1996} model produce an energy spectrum that scales as $E(k) \sim k^{-5/3-2\alpha}$.  This implies that the velocity and energy scalings should become steeper as the sonic Mach number (i.e., degree of compressibiity) increases. 

In a similar vein, \cite{Kritsuk07a} proposed to understand the effects of compressibility on turbulence scalings via the study of the density-weighted velocity (u) power spectrum or structure functions, i.e., $u \equiv \rho^{1/3} v$.  By weighting velocity by the density fluctuations they found that the Kolmogorov scalings can be restored in compressible hydrodynamic turbulence simulations. 
The  \citet{Fleck1996}  and \citet{Kritsuk07a} analytic scalings for compressible turbulence with shocks using hierarchical arguments which were later supported by high sonic Mach number numerical simulations \citep{Kowal07,federrath12,Collins12a,padoan2017ApJ...840...48P}.

The establishment of the GS95 scaling relations and their extension to compressible MHD turbulence  has found many additional astrophysical applications \citep{Vazquez-Semadeni1994,Lithwick2001,CLV03,Padoan04c,Kowal2007,Kritsuk11b}. For instance, the scaling of compressible turbulence has dramatic consequences for understanding cosmic ray propagation and acceleration \citep{Chandran00,YL02,Brunetti_Laz,LY14} as well as for understanding of dust dynamics \citep{HL09,LYgrainrev}.  The effects of compressibility will be discussed in Section \ref{fig:stats} in more detail, as they have great consequences for the statistics of ISM turbulence.

\subsection{Important Parameters of MHD Turbulence in Astrophysical Environments}

It is important to define some basic dimensionless parameters relevant to MHD turbulence.  Such parameters would (ideally) measure the flow properties and describe some universal self-similar behavior of the turbulent medium.   They would therefore be amenable to statistical measurement and characterization.  

\begin{figure}
\includegraphics[width=8.5cm]{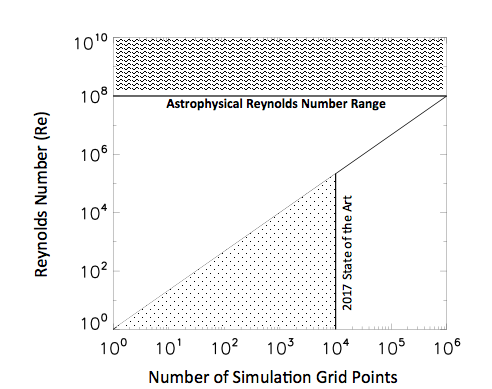}
\caption{Scaling of the number of grid points on a side of a cubic volume vs. Reynolds number. The solid line represents the current state-of-the-art (as of 2017-2021) largest MHD turbulence simulation \citep{2016arXiv160700630F,2020ApJ...904..160F,2021NatAs.tmp...13F}. } This demonstrates the current progress and limitations of computer simulations in reaching the Reynolds numbers of the ISM, which can be larger than Re$=10^{10}$. 
\label{fig:re}

\end{figure}

The most basic dimensionless number of turbulence, the Reynolds number, was previously described in the introduction.  The Reynolds number can be estimated for the ISM from measurements of the size scale of the system (possible if the distance to the object is known), the LOS non-thermal velocity (available from spectral line observations) and the gas viscosity (dependent on the tracer/phase measured).  The Reynolds number in the ISM is much larger than what can be reproduced by numerical simulations (see Figure \ref{fig:re}).  

The ratio of the turbulent velocity to the sound speed, known as the sonic Mach number, characterizes the compressibility of the medium. 
The sonic Mach number is defined as $M_s \equiv V_L/\cs$ \citep[e.g.,][]{Krumholz2005,Federrath2008,Burkhart09}; it thus depends on the turbulent velocity and the sound speed ($\cs$).

There are three primary parameters that encapsulate the importance of magnetic fields in the interstellar medium:
\begin{itemize}
\item The Alfv\'enic Mach number: the ratio of the turbulent kinetic energy to the magnetic energy, which is defined as $M_A=\frac{V_L}{V_A}$, where $V_L$ is the turbulent velocity and $V_A=\frac{|B|}{\sqrt{4\pi\rho}}$ is the Alfv\'en speed. 
\item The plasma $\beta$: the ratio of the thermal pressure ($P_{\rm thermal}$) to the magnetic pressure ($P_{\rm mag}$), i.e.,  $\beta=\frac{P_{\rm thermal}}{P_{\rm mag}}$.  The plasma $\beta$ can also be defined as $\beta=\frac{2M_A^2}{M_s^2}$.

\item The mass-to-flux ratio: this ratio illustrates the importance of comparing the magnetic energy to the gravitational potential energy.
This ratio is commonly expressed as the ratio of the cloud mass to the magnetic critical
mass, $M_\Phi$, which is the minimum mass that can undergo gravitational collapse in  a magnetically dominated medium.
In terms of the magnetic flux,
$\Phi\equiv B c_A \ell_0^2$,
the magnetic critical mass is
\begin{equation}
M_\Phi=c_\Phi\frac{\Phi}{G^{1/2}}\ ,
\label{eq:mphi}
\end{equation}
where $c_\Phi \approx 0.12$ for a cloud with a flux-to-mass distribution corresponding to a uniform field threading a uniform spherical cloud \citep{Mous76}.
For $c_\Phi=1/2\pi$, the ratio of the mass to the magnetic critical mass is
\begin{equation}
\mu_{\Phi,\,0}\equiv\frac{ M_0}{M_\Phi} \propto \frac{M_A}{\alpha_{\rm vir}^{1/2}}
\end{equation}
This relates the virial parameter ($\alpha_{\rm vir}$) to the Alfv\'enic Mach number and magnetic critical mass.
The ratio
$\mu_{\Phi,0}$ is sometimes written as the ratio of the observed
mass-to-flux ratio to the critical one: $(M/\Phi)_{\rm obs}/(M/\Phi)_{\rm crit}$ 
\citep[e.g.,][]{Troland08a,Crutcher2009,McKee2010,Crutcher12a}.
The virial parameter is related to the ratio of the turbulent kinetic energy to the gravitational potential energy
and is defined as:
\begin{equation}
\alpha_{\rm vir}=\frac{5v_L^2R}{GM}
\end{equation}

Gravitationally bound clouds that are both
magnetized and turbulent have $\mu_{\Phi,\,0}$ somewhat
greater than unity (i.e., they are \textit{super-critical}) since the gravity has to overcome both the turbulent motions and the magnetic field \citep{McKee89a,LEC12}. If the cloud is \textit{sub-critical}, then the cloud cannot collapse under ideal MHD due to the frozen-in condition, and therefore a diffusion effect must be invoked, e.g., ambipolar diffusion or reconnection diffusion \citep{Zweibel1983,Mck10,LEC12}.

\end{itemize}

In this review, we will focus on statistics that characterize the turbulence power spectrum (both velocity and density), sonic Mach number, Alfv\'en Mach number, and plasma $\beta$.  If the virial parameter is measured, then the mass-to-flux ratio can also be recovered, although in practice this is not without difficulty \citep{goodman09}. 
Recovering these parameters from astrophysical observations is of enormous importance for star formation, cosmic ray acceleration, and cosmological foreground subtraction and would allow us to characterize many properties of the ISM.  The challenge here is that most of these parameters are difficult to  measure directly from LOS observations, and therefore statistical tools become vital in this pursuit.

\subsection{Direct Numerical Simulations: Promises and Pitfalls}
\label{sec:nums}

While analytic models of turbulence have been successful in predicting basic scaling laws, they are highly idealized in nature. Realistic turbulent flows are frequently inhomogeneous and may not obey the simple statistical scaling discussed above. Therefore, numerical simulations can provide more realistic initial conditions for turbulent flows, as well as taking into account heating and the complexity of the equation of state of fluids.

Despite the successes of numerical simulations of MHD turbulence in reproducing both analytic predictions and observed astrophysical phenomena, a number of major philosophical and practical challenges remain.  In particular, the elephant in the room is the limited physical resolution of numerical simulations, offering at best only an order of magnitude in scale for the turbulence inertial range \citep{Lazssrv09}.  All grid-based numerical simulations are plagued by artificial viscosity, which damps the turbulence cascade in a premature and unphysical way. Smooth-particle hydrodynamic (SPH) simulations are no better off; not only do they have  difficulty capturing shock discontinuities, but accurate treatment of magnetic field dynamics is extremely challenging within this approach \citep{Price10a}.
Perhaps most critically, because of the limited numerical resolution of current MHD turbulence simulations,  these simulations cannot reach the observed Reynolds numbers of the ISM in our galaxy.  Figure \ref{fig:re} illustrates the current progress and limitations of computer simulations in reaching the Reynolds numbers of the ISM, which can be larger than $R_e=10^{10}$.  For grid-based numerical codes, the Reynolds number scales as $N^3 \sim R_e^{9/4}$, where N is the number of grid points.  Current high-resolution MHD simulations can achieve at most $R_e=10^{5}$, running on more than 65,000 cores \citep{2016arXiv160700630F}. Realistic Reynolds numbers will not be reached in the foreseeable future without major advances in our current computational paradigm. 
This implies that we cannot  rely solely on numerical simulations to learn about astrophysical turbulence but must include input and testing from analytic predictions and statistics.

\subsubsection{The Importance of Synthetic Observations}
When combined with statistics, a comparison of numerical simulations with observations can determine the extent to which the numerical ``reality" represents actual physical processes and can provide a path forward on the turbulence problem in astrophysics. In order to compare simulations to observations, the simulations must first be transformed into ``synthetic observations" by choosing a line of sight (LOS) through the numerical data cube and adding telescope beam smoothing, noise and radiative transfer. The output is a column density map, polarization map, integrated intensity map, spectral line, position-position-velocity data cube, or synchrotron fluctuations \citep[see,][for an overview]{2018NewAR..82....1H}. Statistical techniques, such as the Fourier power spectrum,  are then able to extract underlying regularities
of the flow and reject incidental details in both observations and numerical synthetic observations.

\section{Turbulence Statistics}
\label{sec:stats}

\begin{figure*}
\includegraphics[width=16.5cm]{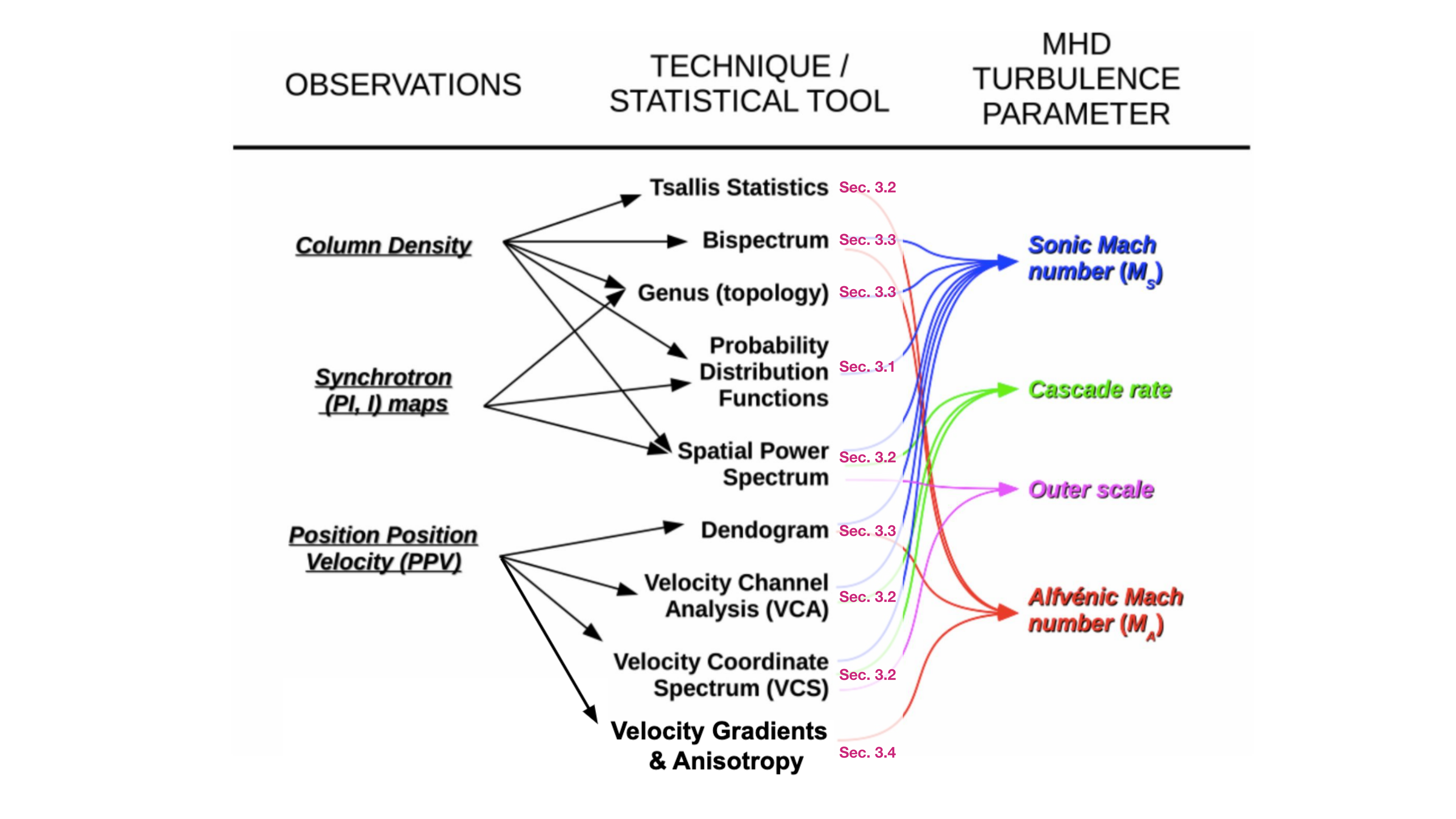}
\caption{
\label{fig:stats}
Observational diagnostics of turbulence  reviewed in this article and their dependence on turbulence parameters of interest such as the sonic and Alfv\'enic Mach numbers, which are the ratios of the flow velocity to the sound speed and the Alfv\'en speed, respectively.
On the far left of the figure is a typical type of observational data, e.g., column density maps, synchrotron maps, and/or radio PPV cubes.  Arrows point to different diagnostic statistics that have been applied and developed for this type of observational data. On the far right, arrows point to the basic turbulence parameter the statistic is designed to inform on. 
Each statistic listed here has been applied to astrophysical data as traced by column density maps, synchrotron maps, and/or radio PPV. For each statistic there is a corresponding section number in this review where we describe its use.   }
\end{figure*}

 In general, the best strategy for studying interstellar
turbulence is to use a synergetic approach, combining theoretical knowledge, numerical
simulations, and observational data. \textit{ This comparison can only be done with the aid of statistics.}
There has been substantial progress in the last decade in the development of statistical techniques to study
turbulence, which is the focus of this review paper. Figure \ref{fig:stats} illustrates how some of these statistics have been shown to depend on the sonic and Alfv\'enic Mach numbers of the fluid flow. 
These techniques have been tested
empirically using parameter studies of numerical simulations and/or with the aid of analytic
predictions.

This review seeks to provide an overview of some common turbulence statistics applied to observational data; however, it will not be possible to cover all statistics and their uses.  Furthermore, we do not go into extensive depth concerning the definitions of  many of these techniques.   The interested reader is encouraged to go to the cited papers to learn more about the implementations and definitions of the diagnostic tools mentioned below.

We have decided to organize the sections that follow based on physical aspects of the turbulent flow.  First we cover diagnostics that target spatial and temporal properties of the energy cascade and compressibility. We organize this by complexity of the statistic, covering one-point statistics (3.1), two-point (3.2), and three-point statistics (3.3.1). Subsequently, we will discuss statistics that can provide insight into turbulent magnetic fields.

\subsection{One-point Statistics: Probability Distribution Functions}
\label{sec:stats_ms}

\subsubsection{Lognormal PDFs}
Perhaps the  most commonly used turbulence diagnostic in the ISM community is the  probability distribution function (PDF). This is in part because isothermal supersonic turbulence is known to produce a lognormal PDF, which lends itself well to analytic analysis \citep{Vazquez-Semadeni1994,Padoan1997,Scalo98a}:

\begin{equation}
p_{LN}(s)=\frac{1}{\sqrt{2\pi\sigs^2}}\exp\left(-\frac{(s-s_0)^2}{2\sigs^2}\right)\,,
\label{eq:LNpdf}
\end{equation}
expressed in terms of the logarithmic density
\begin{equation} \label{eq:s}
s\equiv\ln{(\rho/\meanrho)}\,
\end{equation} where $\sigs$ is the standard deviation of the lognormal.
The quantities $\meanrho$ and $\means$ denote the mean density and mean logarithmic density. 
Deviations from lognormal are expected for different equations of state \citep{federrath2015MNRAS.448.3297F}, but for isothermal turbulence, the width of the lognormal ($\sigs$) is given by the turbulence sonic Mach number $M_s =v_L/\cs$ \citep{Krumholz2005,Federrath2008,Burkhart09,Konstandin12a,Molina12a}, which depends on the rms velocity dispersion ($v_L$), the sound speed ($\cs$) and turbulence driving parameter ($b$):
\begin{equation}
\sigs^2=\ln[1+b^2M_s^2]
\label{eqn.sigma}
\end{equation}

Observations of \textit{column density} from dust or line tracers also suggest a lognormal PDF distribution \citep{berkhuijsen08,Hill2008,burkhart10,Burkhart2015,Imara2016}, although robustly determining the width of the lognormal from observations of dust is generally not possible with current observations \citep{Lombardi2015AA,Alves2017AA,Chen2018}. 
The relationship between the width of the lognormal and the isothermal sonic Mach number takes on a similar form to that of Equation \ref{eqn.sigma}; however, there is a scaling factor, which is proportional to the number of density fluctuations along the LOS.
The process of density perturbation of the column density 
(N$_x$) is similar to the perturbations
of density (n) described above, but the fluctuation occurs only on some fraction of the LOS and therefore each individual fluctuation induces a perturbation
with a smaller relative amplitude. As a result, the dispersion of the PDF will be smaller for column density than for the 3D density distribution. \citet{VazquezSemadeni2001}
proposed the parameterization for determining the form
of the column density PDF as the ratio of the cloud size (or drive scale) to the
decorrelation length of the density field.  \citet{Kowal2007} and \citet{VazquezSemadeni2001} defined the decorrelation length as the lag
at which the density autocorrelation function (ACF) has decayed to 10\% of its original level \citep{Bialy2017ApJ...843...92B,Bialy2020}.

 In molecular clouds with active star formation feedback there is no lognormal distribution in diffuse molecular gas as stellar winds and jets disrupt the low density gas as shown in \citet{Gallegos-Garcia:2020:ApJL:} and Appel et al. (2021).
For the dense gas, recent observational and numerical work has  found that gas in molecular clouds has a power-law PDF rather than a lognormal form \citep{Kritsuk11b,Collins12a,Girichidis2014,MyersP2015,Lombardi2015AA,Burkhart2017ApJ...834L...1B,Mocz2017,Chen2017,Alves2017AA}.  
Observational studies, in particular,  have confirmed that the highest column
density regime (corresponding to visual extinction A$_V >1$) of the PDF has a power-law distribution while the lower column density material in the PDF (traced by diffuse molecular and atomic gas) is described by a lognormal or undefined PDF shape \citep{Kainulainen09a,schneider2015MNRAS.453L..41S,Lombardi2015AA,Burkhart2015,Imara2016,Kainulainen2017}.

While knowing the exact form of the density and column density PDF has many uses, perhaps one of the most important applications is in estimating star formation rates and efficiencies. Most analytic star formation theories rely on supersonic MHD turbulence to produce gravitationally unstable density fluctuations and to set the dense gas fraction \citep{Krumholz2005,Padoan11b,Hennebelle11b,federrath12,Burkhart2018,BurkhartMocz2019}.  These works calculate the SFR by integrating a lognormal or lognormal plus power-law PDF from a critical density for collapse, which usually also varies depending on the parameters of the turbulence and the mean density of the cloud.

\subsubsection{Higher-Order PDF Statistical Moments}

Supersonic turbulence found in the cold gas regions of the ISM of galaxies displays density and magnetic field distributions that are highly non-Gaussian.  As such, higher-order PDF moments, such as skewness and kurtosis, can be useful for describing the statistics of MHD fields, both in 3D simulations and in observations. 

From the point of view of density statistics, \citet{Kowal07a} investigated how the skewness and kurtosis, the 3rd- and 4th-order moments,
respectively, depend on the sonic Mach number. They found that Gaussian asymmetry of density and column density increases as the sonic Mach number increases.  As a result, the PDF moments increase with sonic Mach number. This result was confirmed by later studies with column density maps \citep{Burkhart09} as well as other density-dependent diagnostics such as the rotation measure \citep{Gaensler2011,Burkhart2012,2017AJ....153..163M,Herron2017}.

Observational studies have measured skewness and kurtosis for column density, velocity centroid, and line widths \citep[e.g.,][]{Padoan99a,burkhart10,2013MNRAS.429.1596P,2015MNRAS.446.3777B,2017AJ....153..163M}
Additionally, PDF moments may be applied to individual sub-regions or studied using a rolling circular filter to create spatial moment maps that may be used to study the sonic Mach number and/or areas of local non-Gaussianity \citep{burkhart10,2017AJ....153..163M}.

\subsection{Two-point statistics: Observational Diagnostics for the Turbulence Cascade}
\label{sec:stats_cas}

\subsubsection{Tsallis Distribution: Spatially Dependent PDFs}

The Tsallis distribution is a functional fit to a two-point incremental PDF and \citep{1988JSP....52..479T} was first used as an extension of 
Boltzmann-Gibbs mechanics to multifractal systems.  As MHD turbulence displays characterisitics of fractal systems, the Tsallis functional form was later suggested as a tool for quantifying variations in high Reynolds number flows.  Theses early applications of Tallis to MHD turbulence were done in the solar wind community by  \citet{2004GeoRL..3116807B}.
They investigated temporal variation
in PDFs of magnetic field strength and velocity of the solar wind. The Tsallis statistic is a two-point diagnostic and therefore is highly amenable to applications to time series data such as the solar wind. 

The Tsallis function of an arbitrary incremental PDF $(\Delta f)$ has the form:

\begin{equation}
\label{(tsal)}
R_{q}= a \left[1+(q-1) \frac{\Delta f(r)^2}{w^2} \right]^{-1/(q-1)}
\end{equation}

where $\Delta f(r)=(f(r)-\langle f(r)\rangle_{\bf{x}})/\sigma_{f}$ is the incremental PDF (of some field, e.g., density or velocity or magnetic field, etc.). $r$ is the lag or spatial scale.

The fit is described by the three dependent parameters $a$, $q$, and $w$. The
$a$ parameter describes the amplitude while $w$ is related to the width
or dispersion of the distribution. Parameter $q$, referred to as the ``non-extensivity
parameter'' or ``entropic index'', describes the sharpness and tail size of the
distribution.  
The Tsallis fit parameters are similar to statistical moments.

As a tool for characterizing turbulence in the ISM, studies by \citet{2010ApJ...710..125E} and \citet{Toff11} used the Tsallis distribution to characterize spatial variations in PDFs of 
 density, velocity, and magnetic field of
 MHD simulations. 
 Both efforts
found that the Tsallis distribution provided adequate fits to their PDFs 
and gave insight into statistics of turbulence through the Tsallis fit parameters.
The fit parameters are good diagnostics of the sonic and Alfv\'en Mach numbers. For three-dimensional density, column density, and Position-Position-Velocity data,  \citet{Toff11} found that the amplitude and width of the PDFs (i.e. the w and a parameters of Equation \ref{(tsal)}) show a dependency on the sonic Mach number. They also found that the width parameter is sensitive to the global Alfv\'enic Mach number especially in cases where the sonic number is high. These dependencies held for mock observations, where cloud-like boundary conditions, smoothing, and noise are included.  
Additionally, the Tsallis distribution and the 
Tsallis parameters also can detect the anisotropy of turbulence produced
by the magnetic field \citep{2018MNRAS.475.3324G}.

 \subsubsection{Power Spectrum}

\begin{figure}
\includegraphics[width=8.5cm]{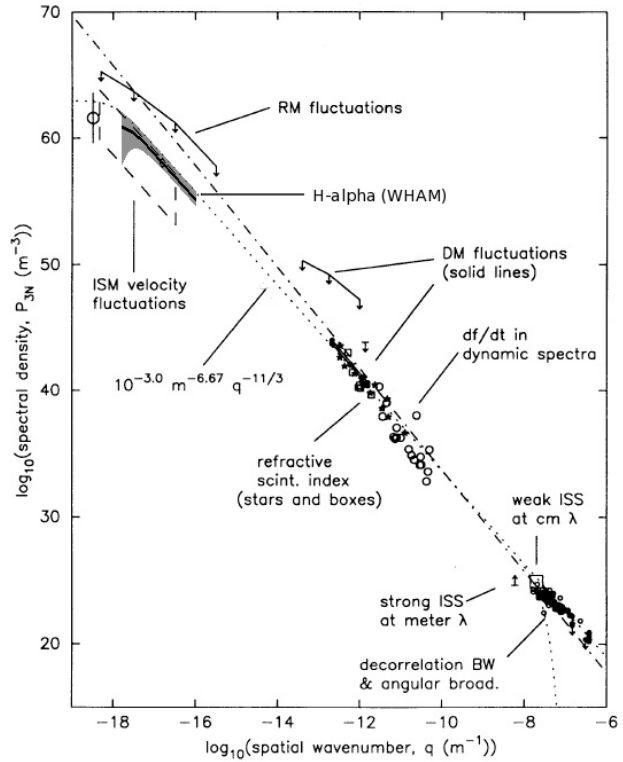}
\caption{The electron density power spectrum, i.e., ``Big Power Law in the sky,'' which extends over 12 orders of magnitude of scale in the galaxy, first reported by \citep{Armstrong95} and later extended by \citep{CheL10} to include data from the Wisconsin H-$\alpha$ Mapper (WHAM).} From \citet{CheL10}. 
\label{fig:bigpl}

\end{figure}

Turbulent motions are self-similar; this gives rise to the power-law behavior seen in the Fourier power spectra and,  conversely, the real-space two-point correlation function. For this reason, the fundamental concept of turbulence as an energy cascade that transfers energy from large scales to small scales is often linked to the Fourier power spectrum. Therefore, the Fourier power spectrum has become a common statistical tool for studying turbulence, including astrophysical turbulence, from observations. For example, the \citet{Kolmogorov41a}  scaling has been observed in the Milky Way galaxy electron density fluctuations in the diffuse warm ISM and is known as the ``Big Power Law in the sky'' \citep{Armstrong95,CheL10}; this is shown in Figure \ref{fig:bigpl}. The  \citet{Kolmogorov41a} scaling seen in electron density fluctuations stretches an impressive 12 orders of magnitude in scale within the Milky Way. 

The power spectrum measures the turbulence energy cascade as a function of scale or frequency and can reveal the sources (i.e., injection scales) and sinks (i.e., dissipation scales) of energy and the self-similar behavior of the inertial range scaling. 
Additionally, the power spectrum can shed light on the spatial and kinematic scaling of turbulence and the sonic Mach
number \citep{LP04,LP06, Kowal07,Heyer09a,burkhart10,Collins12a,2012ApJ...754...29Z, burkhart14,chepurnov15}.

The power spectrum is defined as: 

\begin{equation}
P(k)=\sum_{k}\tilde{F}(k)\cdot\tilde{F}^{*}(k)
\label{eq:ps}
\end{equation}
where $k$ is the wavenumber and $\tilde{F}(k)$ is the Fourier transform of the field under study, e.g., density, kinetic energy, magnetic energy, etc.
One quantity that is related to the Fourier power spectrum is its real space counterpart, the 2D correlation function. The inverse transform that gets the autocorrelation from the power spectrum is defined as:
$$
\Gamma(r)=\int_{-\infty}^{\infty} e^{-i k r} P(k) d k
$$

Additionally, the two-point correlation function and power spectrum are related to the second-order structure function of a field $f(x)$, defined as: 
\begin{equation}
S_{2}(x, r)=\left\langle(f(x+r)-f(x))^{2}\right\rangle
\label{eq:structure_funct}
\end{equation}
It can be related to the power spectrum as:
\begin{equation}
\begin{aligned}
S_{2}=2 \int_{-\infty}^{\infty}\left(1-e^{-i k r}\right) P(k) d k \\
\end{aligned}
\end{equation}

The most direct measurements of the power spectrum of kinetic and magnetic energy in the very local ISM are taken in situ from the Voyager spacecraft \citep{2020ApJ...900..166Z}. Otherwise is not possible to directly measure the kinetic energy power spectrum in the ISM. Furthermore, while inverse cascades can exist in nature \citep{jupiterdyn} and are particularly relevant for the evolution of large scale magnetic fields \citep{Brandenburg2005,PhysRevLett.114.075001,2018PhRvL.120v4502B}, it is not possible to deduce the direction of the energy transfer from ISM observations.  
The vast majority of power spectral measurements in the ISM are taken from  column density maps, integrated intensity maps or velocity centroid maps.  These observables are then related to the underlying 3D density or velocity field of the ISM \citep[e.g.,][]{lp00,esquivel05}, usually with interpretation from simulations.  Similar inferences are also made from measurements of the correlation functions and 2nd-order structure function \citep{2017ApJ...845...53N}.
Traditionally, an observationally motivated power spectrum analysis has involved averaging using mean values, azimuth averaging or averaging along a preferred direction in order to capture anisotropic effects \citep[e.g.][]{Stanimirovic2001,Dickey2001,2003ApJ...583..308P,2015ApJ...809..153M,Kalberla2016A&A...595A..37K,2017ApJ...850...19G}.  A discussion of different averaging procedures for the spatial power spectrum applied to observational column density images is provided in \citet{Pingel2018}.

\textbf{Density Power Spectrum}

The power spectrum (PS) of density can be inferred from column density observations in the optically thin limit \citep{Burkhart2013b}.
For flows with sonic Mach numbers less than unity, the density power spectrum closely follows the velocity power spectrum and is essentially close to the Kolmogorov slope (e.g., the case with the warm media traced by density scintillation and scattering in \citet{Armstrong95}). Additionally, two-point correlation functions have shown Kolmogorov-like scaling in supernova remnants \citep{2020ApJ...894..108R}.

\begin{figure}
\includegraphics[width=8.5cm]{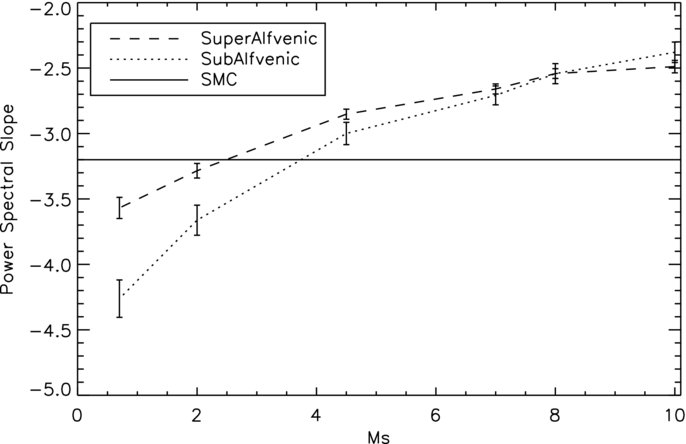}
\caption{
\label{fig:ms}
Reproduced from \citet{burkhart10}: Power spectral slope vs. sonic Mach number for a sub-Alfv\'enic (dotted line) and  super-Alfv\'enic (dashed line) simulation. The spectral slope of the Small Magellanic Cloud (-3.3) is shown as a straight line. The y-axis plots 3D slopes, where 3.667 corresponds to the 1D Kolmogorov slope of 1.667.}
\end{figure}

In the presence of supersonic turbulence, the density
spectral slope is shallower than Kolmogorov because shocks create small-scale enhancements of
density \citep{Kowal07a,Kritsuk07a}. This effect extends to observables, such as column density and/or integrated intensity. 
For example, \citet{burkhart10} plotted the power spectral slope
of column density maps versus sonic Mach number and
found that the slope of the power spectrum of ideal MHD
turbulence is increasingly shallow as the Mach number
increases (see Figure \ref{fig:ms}).  \citet{Collins12a}, \citet{Federrath13a} and \citet{burkhartcollinslaz2015} found that this effect is exaggerated even further in the presence of self-gravity \citep[see also][]{Ossenkopf01a}.
 In some cases,
self-gravitating supersonic turbulence produces density
structures that drive the spectral slope toward increasingly shallow slopes and even, in some cases, positive
slope values. This is in contrast to non-self-gravitating turbulence, where the power is dominated by large-scale structures and the power on the smaller scales is decreasing.
These theoretical expectations were confirmed observationally in a multiphase study of the Perseus molecular cloud by \citet{Pingel2018} and indicate that the column density slope is sensitive to the compressibility and gravitational state of the media. However, it is very difficult to measure the driving and dissipation scales using column density information alone \citep{Yoo2014}.

\begin{table*}
\centering 
\caption{
 Table 2: Studies using VCA/VCS and density power spectrum (PS) to study turbulence in observations. }
\begin{tabular}{| c |c | c | c | c | c | c |}
\hline
& object name & reference & tracer type & technique & velocity slope   & density slope \\ 
\hline \hline 
1 & Arm & \citet{2006ApJS..165..512K}  & HI& VCA  & -1.8 &  -1.2 \\
2 & SMC & \citet{Stanimirovic2001} & HI& VCA & -1.7 &  -1.4 \\
3 & Perseus & \citet{Pingel2018} & dust & PS & - & -0.7\\ 
4 & Perseus & \citet{Pingel2018} & HI & PS & - & -1.2 \\ 
5 & Perseus & \citet{Pingel2018} & $^{12}$CO & PS &- & -1.1 \\ 
6 & Perseus & \citet{Pingel2018} & $^{13}$CO & PS & -& -0.9 \\ 
7 & Perseus & \citet{Sun06a} & $^{13}$CO & VCA  & -1.7 &  -1.0 \\
8 & Perseus & \citet{2006ApJ...653L.125P} & $^{13}$CO & VCA & -1.8  & -1.0 \\
9 & Galactic & \citet{2016AA...593A...4M} & dust & PS & - & -0.9 \\
10 & Galactic & \citet{2019MNRAS.483.3437C} & HI & VCA & -  &  -1.1 \\
11 & MBM 16 & \citet{Pingel2013} & HI & VCA & - & -1.7 \\
12 & Galactic & \citet{2019MNRAS.483.3437C} & HI & VCA & - & -1.1 \\
13 & L1551 & \citet{swift08} & C$^{18}$O & VCA & -1.7 &  -0.8 \\
14 & G0.253+0.016 & \citet{Rathborne15a} & HCN & VCA  & - &  -1.0 \\
15 & G0.253+0.016 & \citet{Rathborne15a} & HCO$^{+}$ & VCA  & - &  -0.9 \\
16 & G0.253+0.016 & \citet{Rathborne15a} & SiO& VCA & - &  -1.1 \\
17 & Orion Nebula & \citet{2016MNRAS.463.2864A} & [S II] $\lambda$6716&VCA & -1.6 &  -1.0 \\
18 & Orion Nebula & \citet{2016MNRAS.463.2864A} & [S II] $\lambda$6731&VCA  & -1.6 &  -1.0 \\
19 & Orion Nebula & \citet{2016MNRAS.463.2864A} & [N II] $\lambda$6583&VCA  & -1.6 &  -0.6 \\
20 & Orion Nebula & \citet{2016MNRAS.463.2864A} & H$\alpha \lambda$6563&VCA  &- &  -0.8 \\
21 & Orion Nebula & \citet{2016MNRAS.463.2864A} & [O III] $\lambda$5007&VCA  & -1.6 &  -0.8 \\
22 & Orion Nebula & \citet{2016MNRAS.463.2864A} &  [O III] $\lambda$5007 & VCA  & -1.4 &  -0.4 \\
23 & SMC & \citet{chepurnov15}  & HI & VCS  & -1.85 & - \\
24 & High galactic latitude &\citet{Chepurnov2010}  & HI & VCS   & -1.87 & - \\
25 &NGC 1333  &\citet{padoan2009ApJ...707L.153P} &$^{13}$CO & VCS  &-1.85 & - \\
26 &NGC 6334 & \citet{2018ApJ...862...42T} & HCN & VCS & -1.66 & - \\
27 & NGC 6334 & \citet{2018ApJ...862...42T} & HCO$^+$& VCS & -2.01 & - \\
28 & Galactic & \citet{2003ApJ...593..831M} & HI & VCA & - & -1.6\\
29 & NGC 5236 &\citet{2020MNRAS.496.1803N} & HI & PS & - & -1.23 \\
\hline
\hline
\end{tabular}
\end{table*}

One important caveat concerning the use of the column density power spectrum as a tracer of the underlying density power spectrum is that opacity effects can alter the slopes. 
\citet{LP04} and \citet{Burkhart2013b} demonstrated that integrated intensity images of optically thick gas saturate to a universal value of -3 (corresponding to a 1D slope of -1) \footnote{For incompressible turbulence, the Kolmogorov power spectrum is k$^{-11/3}$
and  k$^{-5/3}$ for 3D and 1D, respectively.}, regardless of the type of turbulence, nature of shocks, or presence of gravity. 
Their results were also confirmed in later studies who also showed the importance of temperature fluctuations in the lines \citep{boyden2018ApJ...860..157B}.
In light of this, care must be taken when interpreting optically thick line tracers such as $^{12}$CO. The effects of opacity can be seen in this tracer in Table 2. 
In addition to opacity, phase mixing can also affect the density power spectrum \citep{Pingel2018,2019AA...627A.112K}.

\textbf{Velocity Power Spectrum}

In order to measure the power spectrum of the turbulent velocity field from observations, additional care must be taken.  This is because spectral-line data cubes contain information concerning the emissivity and the line-of-sight velocity together, and their separation is nontrivial.   Additional difficultly in measuring the velocity power spectrum comes from the contribution of thermal broadening and phase transition effects, which can mask the underlying turbulence signal, especially in 21-cm emission line observations, which have significant temperature fluctuations \citep{2019ApJ...874..171C, 2020arXiv200301454K}.
Despite these difficulties, the velocity power spectrum of the ISM remains a coveted measurement since it more directly represents the
dynamics of the turbulent flow as compared to integrated density statistics.  We aim to review key techniques for recovering the velocity power spectrum while at the same time urging caution regarding their application to multiphase tracers.

The problem of disentangling velocity and density fluctuations was first
addressed in \citet{lp00} and in subsequent papers \citep{LP04,LP06,LP08}, as well as in a review article by \citet{Lazarian07rev}.
\citet{lp00} first demonstrated analytically that it is possible to separate the contributions of velocity and density in a radio spectral-line cube (e.g. PPV cube) integrated over spectral frequencies.  
Based on this, there are two complementary techniques that have been developed to divorce the velocity power spectrum from emissivity effects in a radio PPV data cube: the Velocity Channel Analysis technique (VCA)
 and the Velocity Coordinate Spectrum technique (VCS).
Below we provide a
brief introduction to  the mathematical foundations of the VCA and the VCS techniques.

The goal of the VCS and VCA is to relate the two-point statistics of a PPV cube to the underlying \textit{true} turbulence two-point scalings, i.e. the density and velocity power spectral or correlation function scalings.  Thus both techniques begin with writing down the PPV correlation function, 
$\xi_s(R,v)\equiv \langle \rho_s({\bf X_1},v_1)\rho({\bf X_2}, v_2)\rangle\\ \nonumber $ 

 where the LOS-axis (here defined as z) component velocity
$v$ is measured in the direction defined by the two dimensional
vector
${\bf X}$, $R$ is the spatial separation
between points in the plane-of-sky, and $\rho_s$ is the PPV energy density:
\begin{equation}
\rho_s ({\bf X}, v)\sim \int^S_{-S} dz \rho({\bf X}, z) \phi_{vz}({\bf X}, z)
\label{first}
\end{equation}
 
where the emission is coming from a cloud of size $2S$ and the intensity
of emission is  proportional to the cloud density $\rho$. $\phi_{vz}$ is the distribution function of the z-component of velocity that takes a Maxwellian form.

If the gas is confined in an isolated cloud of size $S$, the zero-temperature correlation
function is (see LP06)
\begin{eqnarray}
\xi_s(R,v)\propto
\int_{-S}^S {\mathrm d}z \left(1-\frac{|z|}{S}\right)
\; \frac{\xi( r)}{D_z^{1/2}({\bf r})}
\exp\left[-\frac{v^2}{2 D_z({\bf r})}\right],
\label{ksicloud}
\end{eqnarray}

where $\xi( r)=\langle \rho ({\bf x}) \rho ({\bf x}+{\bf r}) \rangle$ is the correlation function of the density field and, for a turbulent medium, can be written in terms of a 
 part that depends on the mean density
only and a part that changes with $r$, i.e, $\xi(r)= \langle \rho \rangle^2 
\left(1 + \left[ {r_0 \over r} \right]^\gamma\right)$, where  $r_0$ is the scale at which fluctuations
are of the order of the mean density.
$D_z$ is the correlation function of the underlying true turbulent velocity field and (for a power law) can be written as:
$D_z\sim D(L) (r/L)^m$.

The aim now is the relate the properties of the observable in Eq. \ref{ksicloud} to the exponents $m$ and $\gamma$ which given the turbulent velocity and density two-point scaling.
 The key is that  the PPV correlation
function $\xi_s$ can be presented as \textit{a sum of two
terms, one of which does depend on the fluctuations of density, the other does not.}  Hence it is possible to separate density and velocity fluctuation in PPV.

Taking Fourier transform of $\xi_s$ one gets the PPV spectrum $P_s$, which is also
a sum of two terms the power spectrum of density $P_{\rho}$ and  the power spectrum of velocity $P_{v}$, $P_s=P_\rho+P_v~~,$
where the asymptotic (or numerical) solutions for $P_{\rho}$ and $P_{v}$ in one dimension (along the velocity axis) and two dimensions (in the velocity slice) form the basis of VCA and VCS. Functionally,  the difference between the two is that VCA takes a two dimensional power spectrum of a 2D channel slice while the VCS takes a 1D power spectrum along the velocity axis.
Whether $P_\rho$ or $P_v$ dominates depends on the density spectral index $\gamma$. If $\gamma<0$ the $P_\rho$ contribution is always subdominant. The density power spectrum can be found from the column density power spectrum discussed above, atleast in the optically thin limit \citep{Bur13}.  This given an independent measure of $\gamma$ and allows VCS or VCA to contrain $m$. Table 1 and 2 of \citep{Lazarian07rev} give the asymptotic scalings of the VCS/VCA. It is also possible to numerical fit the full VCA/VCS integrals, which was the approach taken in \citet{chepurnov15}
We further describe some resulting applications of these two techniques below. Table 2 shows estimated velocity slopes from application of VCA and VCS as well as some studies using column density power spectral analysis (PS) to get the column density power spectral slope.

\textbf{VCA}:

As discussed above, VCA works by computing the 2D power spectrum over different PPV channel thicknesses. As the spectral channel width increases, the power-spectral index approaches the limit of the power spectrum of the full  column density, which is produced by density fluctuations in the optically thin limit. \citep{Stanimirovic2001,LP06,Burkhart2013b}.
For sufficiently small spectral channel widths, the 2D channel map power spectrum will be increasingly dominated by the contribution of velocity. As the channel width increases, the power spectral index becomes dominated by the density fluctuations \citep[e.g.,][]{MullerBisk,Pad06,2003ApJ...593..831M}. 

A break in the 2D slope as a function of velocity channel thickness indicates a transition from velocity-dominated to density-dominated maps, and the velocity spectrum can be inferred by relating this shift to the analytic predictions outlined in \citet{LP04}.

\textbf{VCS}:

As described previously, the VCS technique takes a 1D spectral power spectrum along the velocity axis over varying spatial beam sizes.  As outlined in   \citet{LP06} and \citet{Chepurnov2010}, there are analytic asymptotic  solutions for the high- and low-resolution limits, which can be fitted  to recover the velocity power spectral slope. It is also possible to directly calculate and fit the 1D spectral line power spectrum and relate this back to the slope of the density power spectrum, the slope of the velocity power spectrum and the injection scale of the turbulence, as was done in \citet{chepurnov15}.
 Further more, VCS also has a unique features which give it an advantage over the VCA. It is applicable to data cubes that are not spatially resolved and can therefore be used on absorption lines. 

\begin{figure}
\includegraphics[width=8.5cm]{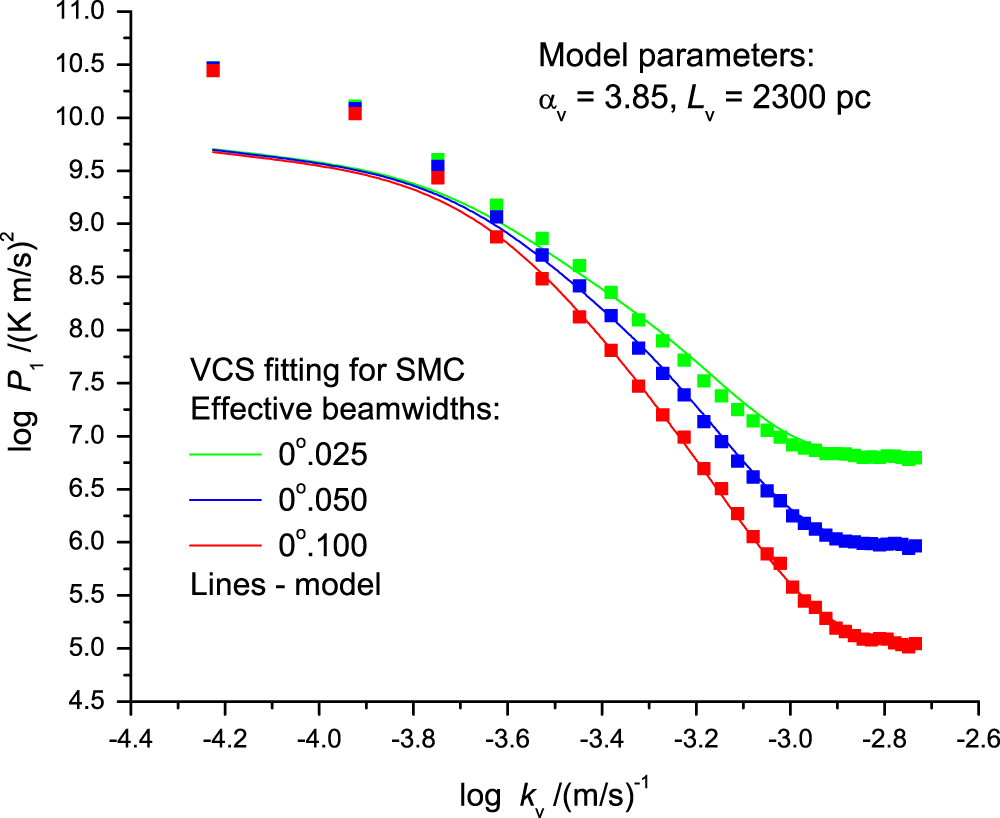}
\caption{
\label{fig:vcssmc}
Reproduced from \citet{chepurnov15}: Fitting of the model VCS  of the Small Magellanic Cloud observed at 21 cm for different
effective resolutions. The fitting to the VCS (Eq. 12 of \citet{chepurnov15}) gives the following turbulence parameters for the Small Magellanic Cloud:
velocity slope of -3.85, density slope of -3.3, and  driving scale of 2.3 kpc.}
\end{figure}

We highlight an example application of the VCS technique from \citet{chepurnov15}.  They used the VCS technique to calculate the velocity power
spectrum in the Small Magellanic Cloud (SMC) observed in 21-cm emission.  We show the example of their VCS fitting in Figure \ref{fig:vcssmc}.  The VCS is able to recover the velocity power spectrum, fitted injection scale and other properties of the turbulence.

For additional details about VCA and VCS see the review paper by \citet{Lazarian07rev}.

\subsubsection{ Wavelet Transforms and Delta-Variance}

A number of wavelet transforms have been used for studies of turbulence. Similar to a Fourier transform or a structure function, a wavelet measures the amount of structure on varying spatial scales by taking the sum of values in a wavelet decomposition \citep{Kowal07a,Gill:1990:ApJL:,Zielinsky1999A&A...347..630Z}. A commonly used example of a wavelet transform technique for turbulence studies is the application of the so-called delta-variance technique \citep{2001A&A...366..636B,Ossenkopf02a}. The delta-variance is an extension of Allan variance for one-dimensional time series. The delta-variance characterizes the structure found in a map at a particular spatial scale by computing the variance in the wavelet decomposition.  It can also be applied to irregularly-shaped observational maps  \citep{2008A&A...485..917O,2008A&A...485..719O}.

Wavelets are also highly adept at picking up non-Gaussian features, which can be produced by shocks. Recently, the wavelet scattering transform (WST), a low-variance statistical description of non-Gaussian processes which was developed in data science and encodes long-range interactions through a hierarchical multiscale approach based on the wavelet transform, has been applied in the context of MHD turbulence \citep{2019A&A...629A.115A}. 
Many applications of the WST have been found in astrophysics, including applications to cosmology and  large-scale structure syntheses \citep{2019A&A...629A.115A,Cheng:2020:arXiv:}. For the ISM, 
the WST and similar reduced forms have been applied to analyze anisotropies in HI from the Canadian Galactic Plane Survey \citep{Taylor:2003:AJ:,2006ApJS..165..512K}   and non-Gaussianity \citep{Robitaille:2014:MNRAS:,Taylor:2003:AJ:} in the Herschel infrared Galactic Plane Survey (Hi-GAL) \citep{Molinari:2010:PASP:}.  Recently, a reduced form of the WST (RWST) was shown to characterize  MHD simulations (in 2D) \citep{2019A&A...629A.115A} and has  subsequently been applied to Herschel observations  as well as to polarized dust emission \citep{Regaldo-SaintBlancard:2020:arXiv:}.

The study by \citet{2019A&A...629A.115A} and \citet{2021ApJ...910..122S}
 shows that the WST can infer the properties of turbulence and produce realistic synthetic fields as well as retaining couplings between scales.  This shows that wavelets and related transforms are extremely promising tools that can be used to help characterize observational turbulence.
As the uses of wavelets in turbulence studies is a vast subject, we direct the interested reader to \citet{2015JPlPh..81f4302F} for a deeper review.

\subsection{Three-point statistics, phase information, and topology}
\label{sec:bispec}

As discussed above, the Fourier power spectrum only contains  amplitude information and rejects all information regarding phases in a dataset. This is problematic because much of the information in an image or 3D field is contained in the phases.  There are several higher-order techniques that preserve phase information and have been applied to turbulence simulations and observations, a subset of which will be reviewed here.

\subsubsection{Bispectrum}
One commonly used Fourier-based technique that preserves phase information is the bispectrum \citep{1989PhFlB...1..271I,2000ApJ...544..597S}. The bispectrum is a higher-order statistic in relation to the power spectrum. The Fourier transform of the second-order
cumulant (the autocorrelation function) is the power
spectrum (see Equation \ref{eq:ps}), while the Fourier transform of the third-order cumulant is known as the bispectrum. 

The bispectrum is defined as:

 \tiny
\begin{equation}
B(|\overline{k_{1}}|,|\overline{k_{2}}|)=\sum_{|\overline{k_{1}}|=const}\sum_{|\overline{k_{2}}|=const}A(|\overline{k_{1}}|)\cdot A(|\overline{k_{2}}|)\cdot {A}^{*}(|\overline{k_{1}}|+|\overline{k_{2}}|)
\label{eq:bispectra}
\end{equation}
\noindent
\normalsize

\begin{figure}
\includegraphics[width=8.5cm]{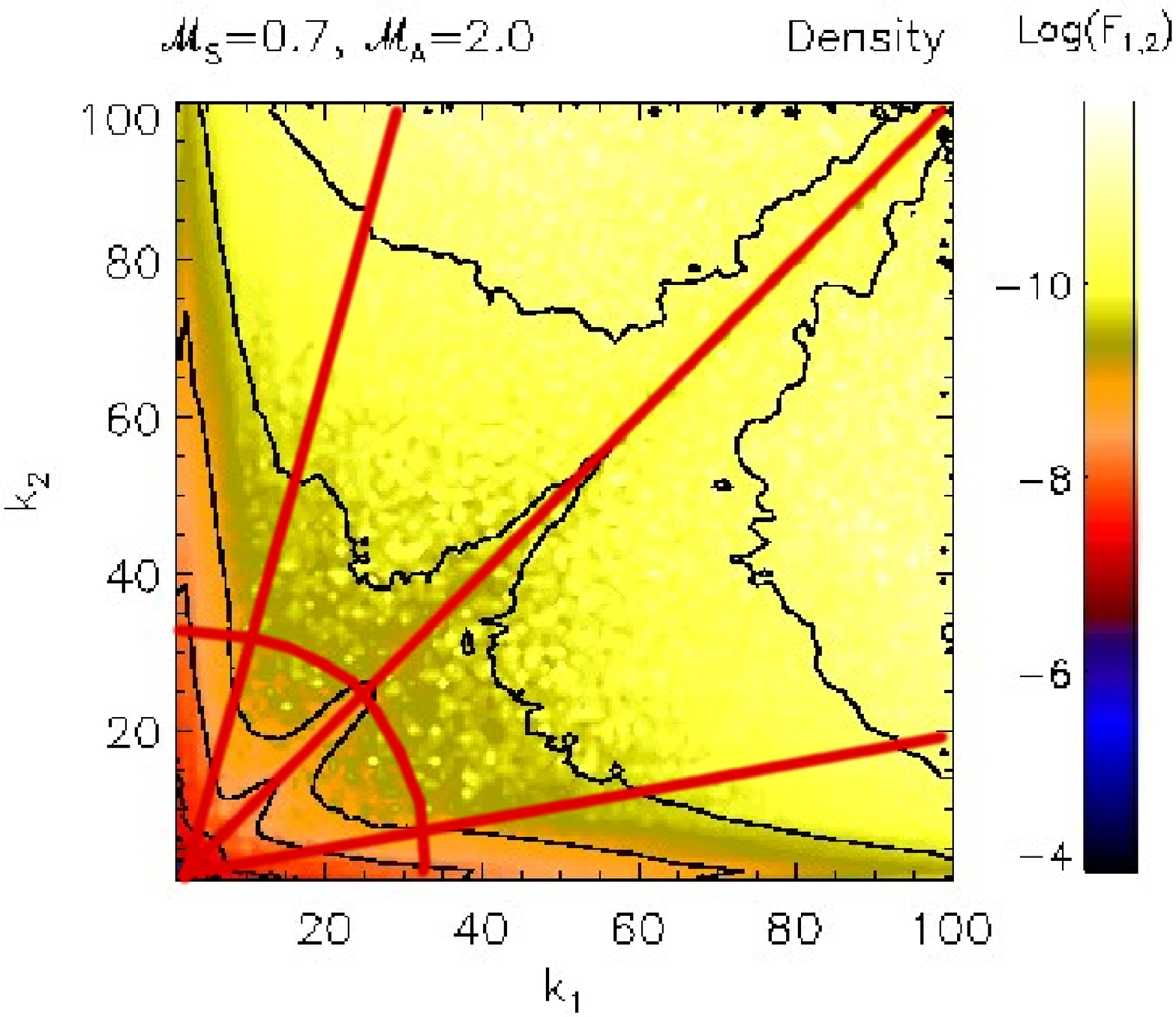}
\includegraphics[width=8.5cm]{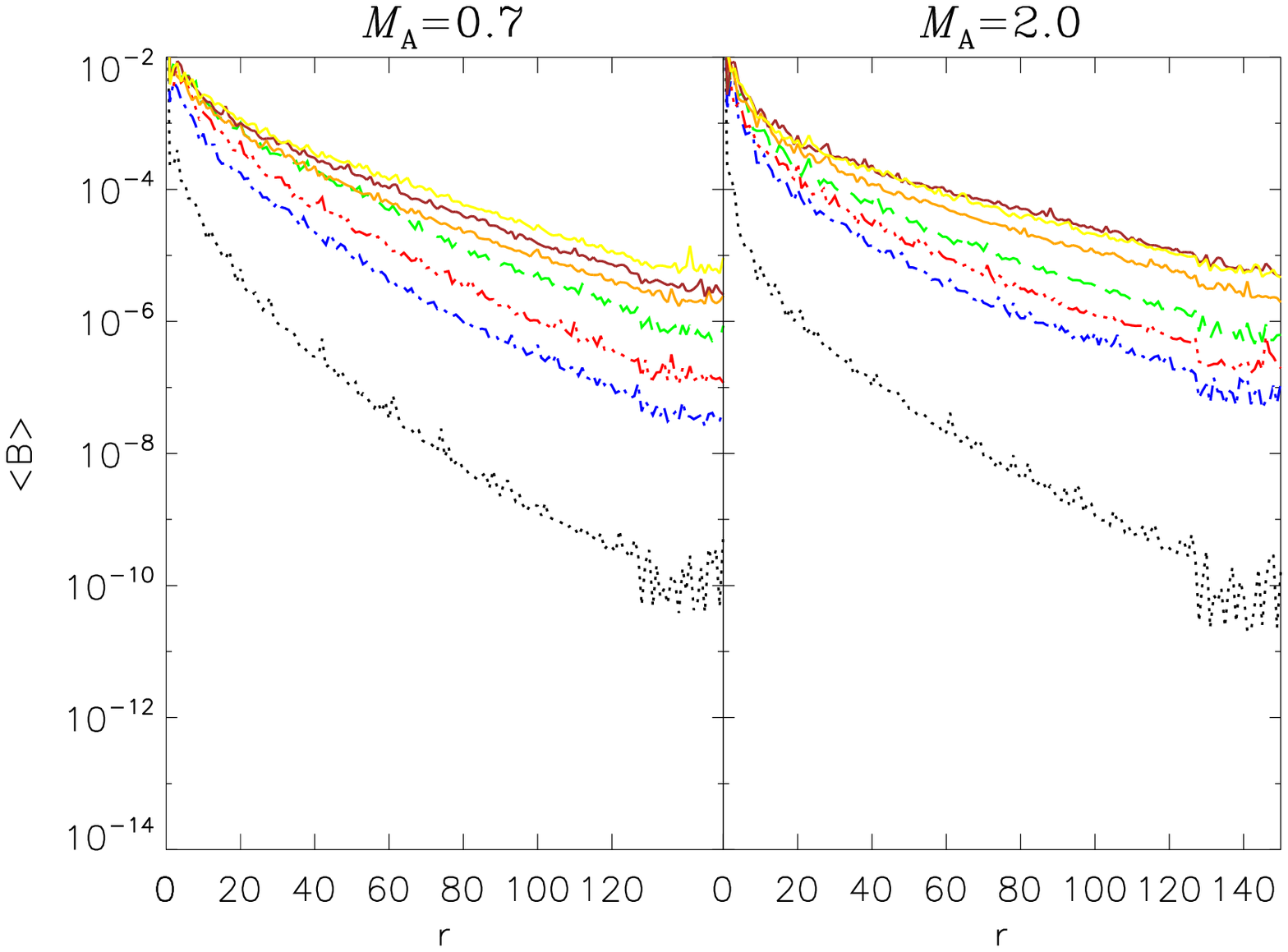}
\caption{
\label{fig:bispe}
Reproduced from \citet{Burkhart2016ApJ...827...26B}: Top: an example of two  bispectral averaging procedures to reduce information: 
One can perform radial averaging over bispectral amplitude values for a given $R^2=k_1^2+k_2^2$ (e.g., the red circular curve that intersects $k_1=k_2$) or angular averaging of the bispectral amplitudes over a given angle  ($\alpha$) as measured from zero (e.g., the three red radial lines).
Bottom: bispectral amplitudes of column density along the magnetic field (X-direction) averaged along different annuli, i.e.,  for a given $r^2=k_1^2+k_2^2$. Different colors indicate different sonic Mach numbers, with black being subsonic and yellow being the most supersonic model. }
\end{figure}

The bispectrum preserves 
 information on both amplitudes and phases and
has been applied to simulations and observations \citep{Burkhart09,Cho2009,burkhart10} in the context of characterizing non-Gaussian features arising from compressible turbulence. These studies found that the bispectrum is a sensitive diagnostic
for both the sonic and the Alfv\'enic Mach numbers.  More importantly, the bispectrum can describe the behavior of nonlinear mode correlation across
spatial scales. While the power spectrum is a two-point statistic averaged over a single wavevector, the bispectrum is three point statistic (formed of triangles) and hence is a function of two wavevectors. Thus, studies involving bispectrum often plot isocontours of the bispectral amplitudes. Figure \ref{fig:bispe} top shows an example bispectral contour map of a MHD turbulence simulation. Higher bispectral amplitudes highlight areas of increased non-linear interactions and deviations from Gaussianity. Reflecting the forward energy cascade observed in these simulations, the bispectral amplitudes decrease towards smaller scales.  Because it is a two dimensional metric, the bispectrum is more difficult to interpret than the power spectrum. To address this issue, \citet{Burkhart2016ApJ...827...26B} introduced averaging over bispectral contours in order to reduce the information
provided by the bispectral amplitudes (see Figure \ref{fig:bispe}). An example of bispectral averaging is shown in the bottom of Figure \ref{fig:bispe}. Two possible averaging procedures are reflected in the top panel of Figure \ref{fig:bispe} in red lines (angle averaging or for a given annuli).  Figure \ref{fig:bispe} bottom shows averaging along different annuli.  Higher bispectral amplitudes are clearly seen in the averaging for higher Mach number turbulence.

The bispectrum technique characterizes and searches for
nonlinear interactions and non-Gaussianity, which makes it a useful
technique for studies of supersonic super-Alfv\'enic MHD turbulence in the ISM and solar wind due to the fact that as turbulence eddies evolve, they transfer energy from 
large scales to small scales, generating a hierarchical turbulence cascade.
For incompressible Kolmogorov-type flows, this can be expressed as $k_1 \approx k_2 = k$ and $k_3 \approx  2k$. 
Nonlinear  interactions take place more strongly in 
compressible and magnetized flows, i.e., $k_1 \ne k_2$.
The bispectrum as well as other three-point statistics can characterize these nonlinear interactions.

Related to the bispectrum is the three-point correlation function (3pcf).  The 3pcf is the Fourier transform of the bispectrum, i.e., it is the real-space analogy of the relationship between the two-point correlation function and the power spectrum, discussed previously.  A fast, angle-dependent multipole expansion was applied to MHD turbulence simulations in  \citet{Portillo:2018:ApJ:,3PCFkeyunpub}. The angle dependency is also important as most implementations of the bispectrum and the 3pcf often average over angles to reduce the amount of information. 
\citet{Portillo:2018:ApJ:} found that the multipole representation of the 3pcf is mildly sensitive to the sonic and Alfv\'enic Mach numbers of turbulence.

Phase information may also be explored via the phase coherence index \citep[(PCI)][]{hada2003SSRv..107..463H,koga2003SSRv..107..495K,2008NPGeo..15..567C,Burkhart2016ApJ...827...26B}. 
The PCI creates a surrogate data set that has phase information randomized and another surrogate data set that 
has its
phase information perfectly correlated.  These surrogate data share the same
power spectrum  and  original data under study,  but have different 
phases.  Comparison of the original data and the surrogates gives insight into the 
level of phase coherence or randomness.  
\citet{Burkhart2016ApJ...827...26B} studied the PCI and  found that the most correlated phases occur in supersonic sub-Alfv\'enic
turbulence and also near the numerical dissipation regime. Their study showed that nonlinear interactions with correlated phases are strongest in shock-dominated regions, in agreement with similar studies on solar wind in-situ measurements.

\subsubsection{ Dendrograms} 

MHD  turbulence is able to create hierarchical structures in the ISM that are correlated on a wide range of scales via the energy cascade. A commonly used measure of hierarchical clustering is a form of tree diagram known as a dendrogram \citep{goodman09,Burkhart2013,2013ApJ...777..173B}.  Dendrograms have been used  to characterize structures in synthetic PPV emission cubes of both observations and simulations.  These structures and their nesting properties are related to the presence and strength of self-gravity and compressible density fluctuations produced by supersonic turbulence.
Dendrograms can be used in combination with other statistics, such as 1D PDFs \citep{Chen2018}, and have been used to study the ratio of kinetic to gravitation energy (the Virial parameter) in molecular clouds \citep{goodman09}.

\subsubsection{Genus Topology Tool } 
The topology of density in the ISM can provide insight into ISM physics and phase structure. For
example, in the \citet{1977ApJ...218..148M} model of the ISM, the topology different phases is expected to be different for gas in the hot phases (hole-like topology) as compared to meat-ball like topology of the warm and cold phases.  This topology is in contrast to other models of the ISM in which the cold and warm gas is more volume filling \citep{1974ApJ...189L.105C}.
Supersonic turbulence also produces drastic changes in gas topology as compared to subsonic turbulence and for this reason, \citet{2002ASPC..276..182L} suggested the use of the Genus statistic \citep{1986ApJ...306..341G} as a measure of ISM topology. 

A  genus statistic is a measure of topology that can distinguish between  hole topology (`swiss cheese') vs. clumpy topology  (`meatball').
The value of the genus is the difference between the number of isolated regions/contours above and below a set threshold value.  The genus curve is produced by varying the threshold value and constructing a curve from the numbers of high and low peaks.  
The genus has extensive use in cosmology and was first applied to interstellar MHD turbulence by
 \citet{Kowal07a}.  Subsequent studies by  \citet{Chepurnov2008} and \citet{2012ApJ...749..145B} applied the genus to the SMC galaxy and to synchrotron polarization data, respectively.  These studies all point to the use of genus in measuring the compressibility of turbulence, since shocks produce changes in the topology of the medium and induce a clumpy topology.

\subsection{Magnetization}
\label{sec:stats_ma}

A number of new techniques have been developed specifically for tracing turbulent magnetic fields and characterizing the Alfv\'enic nature of the interstellar medium \citep[e.g.][]{2017ApJ...850....4H,2018A&A...614A.101V}.   As described in Section \ref{sec:theory}, MHD turbulence closely links the behavior of the magnetic field with the overall state of the gas via scale-dependent anisotropy and the alignment of turbulence eddies with the magnetic field \citep{GS95,LV99,2001ApJ...554.1175M,Cho2002,Brandenburg2005}.
In this subsection we review three such techniques that are designed to trace the Alfv\'en Mach number in a statistical sense in the molecular and neutral ISM.

\subsubsection{Histogram of Relative Orientation}

Recently, observations made by \textit{Planck Telescope} \citep{Planck2016}, HERSCHEL \citep{2013A&A...550A..38P},  and BLASTpol \citep{2017A&A...603A..64S}, have revealed the relation between the filamentary structures and the magnetic field of these regions. These observations show that dense structures usually appear perpendicular to the magnetic field \citep{2013A&A...550A..38P} while less dense structures appear mainly aligned to the projected magnetic field in the plane of the sky \citep[see][]{2016AA...586A138P,2019A&A...629A..96S}. A commonly used method for tracing the change in orientation of the magnetic field relative to dense ISM structure is the Histogram of Relative Orientation (HRO), introduced for use on polarization data in \citet{So13}.
The HRO utilizes  gradients to characterize the statistical directionality of density and column density structures on multiple scales relative to the magnetic field orientation. It can be applied to observations (using gradients of column density) or in 3D using numerical simulations. A summary of the application of the HRO and source code can be found as part of the MAGNETAR project \citep{solar2020}. MAGNETAR calculates the histogram of relative orientation between density structure in the magnetic field in data cubes from simulations of MHD turbulence and observations of polarization using the method of histogram of relative orientations (HRO). 

Application of the HRO to 3D turbulence simulations confirms that dense filaments in molecular clouds are either structures that are  produced by shocks \citep{Hennebelle2013A&A...556A.153H,Federrath2015} or shock-induced self-gravitating \citep{mocz2018MNRAS.480.3916M} features that pull the magnetic field to a perpendicular angle relative to the density gradient \citep{So13,soler2017A&A...607A...2S}.  This transition to perpendicular orientation happens only in the presence of trans- or sub-Alfv\'enic turbulence, since in strongly super-Alfv\'enic turbulence the magnetic field is randomly oriented with respect to density \citep{2021MNRAS.503.5425B}.
 This is evidence that molecular clouds on the 10pc scales are probably at least trans-Alfv\'enic. \cite{2020arXiv200300017S} and \citet{2021MNRAS.503.5425B} also point out that the ability to observe the HRO transition from structures parallel  to the magnetic field to structures perpendicular to the magnetic field is dependent on the line of sight (LOS) relative to the mean magnetic field. 

\begin{figure}
\includegraphics[width=8.5cm]{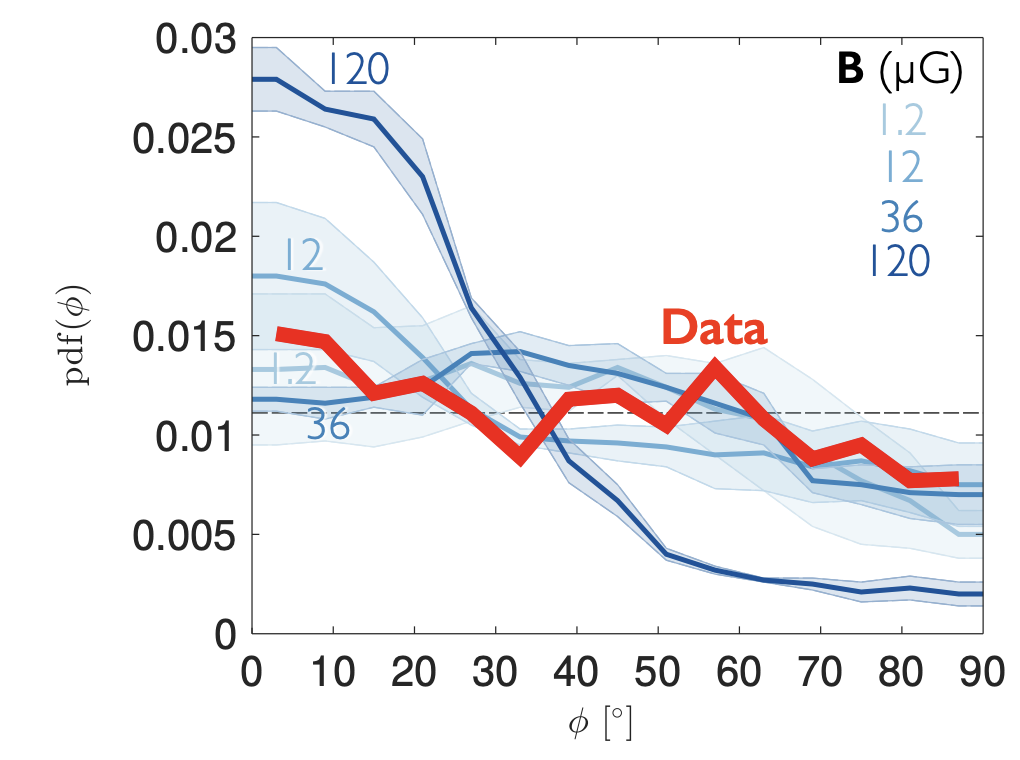}
\caption{
\label{fig:hro}
Reproduced from \citet{2017ApJ...842L...9H}: Histogram of Relative Orientation: shown are
the calculated HRO relations between the magnetic field orientation
and the density gradients in the Ser-emb 8 ALMA data (red line), as well
as in AREPO MHD turbulence simulations with gravity sampled in a 3000 AU region. The ALMA data of Ser-emb 8
show randomly distributed orientations, consistent with the simulations with M$_A> 1$. The shaded error regions
reflect the variation in the multiple lines of sight used to calculate the HRO in the simulations.}
\end{figure}

At smaller scales (i.e., down to 1000 AU scales), \cite{2017ApJ...842L...9H} and \citet{Mocz2017} studied the HRO and morphology of the magnetic field around collapsed cores using ALMA data of Ser-emb 8 and 3D MHD simulations (see Figure \ref{fig:hro}). 
They found that the magnetic field in the Ser-emb 8 core deviates  from the
ideal theoretically motivated collapsing, non-turbulent,
magnetically supported core that has an hourglass-shaped
field.  The HRO shapes suggest that modern analytic models of star formation must take into account turbulent initial conditions.
The random field morphology and the lack of an hourglass
shape are evidence that the magnetic field in Ser-emb 8
is dynamically unimportant and that trans-Alfv\'enic conditions may have been present in the initial configuration of this star-forming cloud.
However, an hourglass-shaped field at the 1000 AU scale is also observed in the strongly magnetized simulations.  Systems that do exhibit hourglass morphology may have initially sub-Alfv\'enic turbulent states.

\subsubsection{Magnetic Anisotropy}

Several different techniques have been tuned to be sensitive to large-scale anisotropy since the first discussions of observational measurement of
the large-scale anisotropy in \citet{2002ASPC..276..182L}.

\textbf{PCA}:

One of the earliest techniques proposed for gauging the degree of anisotropy in the ISM is the Principal Component Analysis (PCA).  PCA is a general dimensionality reduction statistic with many applications that 
searches for correlated components based on the decomposition of the data's covariance matrix. 
\citet{Heyer1997ApJ...475..173H}  were the first to suggest using PCA to search for magnetically induced anisotropy in the ISM.  It has since been applied for this purpose in  both simulations and observations and was further developed and applied in several subsequent works and applied to observations \citep{2002ApJ...566..276B,Brunt09a,2013MNRAS.433..117B,2011ApJ...740..120R,2008ApJ...680..420H}. 

\textbf{Two Point Statistics}:

Structure functions and correlation functions have also been employed to search for anisotropy in observations. \citet{2003MNRAS.342..325E} proposed to measure contours of equal correlation corresponding to different velocity
channel thicknesses in HI and CO data (i.e., a procedure similar to the Velocity Channel Analysis discussed previously). The amount of anisotropy can be linked to the overall strength of the magnetic field.  Structure functions of velocity centroid maps can demonstrate significantly elongated isocontours, which is indicative of anisotropy induced by a strong mean magnetic field \citep{esquivel11,2020MNRAS.496.4546H}. However, this technique suffers from LOS confusion and the orientation relative to the mean field \citep{burkhart14}.

\subsubsection{Velocity Gradients}

In addition to overall anisotropy,  MHD theory predicts that the velocity fluctuations trace magnetic field fluctuations surrounding the turbulent eddies.   An observational consequence of this is that the amplitude of velocity gradients across eddies increases with decreasing spatial scales. As a result,  velocity gradients can be used as a tracer of magnetic fields on small scales, also where anisotropy is present.

The Velocity Gradient Technique (VGT)  was first introduced in \citet{casanova2017ApJ...835...41G} to be applied to radio PPV data. This method was further quantified and extended in \citet{Yuen2017ApJ...837L..24Y}. A recent detailed comparison with other statistical techniques for measuring magnetic fields (e.g., HRO and PCA) in the interstellar medium was presented in \citet{yuen2018ApJ...865...54Y} and \citet{2020MNRAS.496.4546H}. \cite{2018ApJ...865...46L} showed that the VGT is a sensitive probe of different Alfv\'en Mach numbers 
and can also trace the overall direction of the magnetic field, as seen in polarization data from BLAST \citep{2013A&A...550A..38P,2017A&A...603A..64S} and Planck \citep{2016AA...586A138P}.  Additional studies suggest that gradients can also be used for tracing the sonic Mach number and for finding regions dominated by gravitational collapse in molecular clouds \citep{2020ApJ...898...65Y,2020ApJ...897..123H}.
The results of these recent velocity gradient studies have demonstrated the  ability of gradients to trace magnetic fields and determine the 3D magnetic field strength.

Another promising use of gradients for tracing turbulent magnetic fields is their use with synchrotron data. We refer the interested reader to recent studies for a more in-depth discussion \citep[e.g.][]{2018ApJ...863..197Z,2019ApJ...887..258H,zhang2020ApJ...895...20Z}

\section{Discussion: A Path Forward for the Turbulence Problem by Combining Numerics, Observations, Machine Learning, and Statistics}
\label{sec:pathforward}

Despite the near-future inability of numerical simulations to reach the actual Reynolds numbers of astrophysical environments, our evolving understanding of turbulence suggests that simulations are able to reproduce the statistical features of the MHD turbulence cascade and related physical processes.
What aspects or range of scales in MHD numerical simulations can one believe is representative of physical reality and what part is numerical artifact?   There are three ranges of scales of interest in a turbulent energy cascade: the energy injection at large scales ($k_{\rm{inj}}$); the dissipation of energy at smaller scales ($k_{\rm{diss}}$); and the range of scales where the cascade is self-similar (the inertial range), which exhibits power-law behavior in the Fourier power spectrum and is between the injection and dissipation scales.
The behavior of turbulence in the vicinity of  $k_{\rm{inj}}$ and $k_{\rm{diss}}$  is not universal and depends on the details of the numerical setup.  In all current numerical simulations of turbulence, the separation between the energy injection scale and the energy dissipation scale is many orders of magnitude less than in nature.  Nevertheless, if this separation is sufficient\footnote{What separation range is sufficient depends on the nature of the cascade, and in particular, on the range of scales in which the eddies influence neighboring eddies. For instance, to get the correct  slope, this separation should be larger for MHD turbulence than for hydrodynamic turbulence \citep{BersLaz2010}.}, then numerical simulations can deliver the correct statistical properties of the turbulent cascade, e.g., the turbulence Fourier power spectrum, despite the low Reynolds numbers. The power spectral slope can then be compared with and tested against observations.

In recent years, simulations have been best compared with observational data via the creation of ``synthetic observations."  These can include choosing a line-of-sight (LOS) through the numerical data cube, adding telescope beam smoothing or noise, and transforming the simulation in a radio position-position-velocity (PPV) cube or column density map. At the same time, statistical techniques are able to extract underlying regularities
of the flow and reject incidental details in both observations and numerical synthetic observations.

\subsection{Synergistic Comparison of Numerical Simulations, Observations and Theory}

When combined with statistics, a comparison of numerical simulations with observations can determine the extent to which the numerical ``reality" represents actual physical processes. 
In general, the best strategy for studying a difficult subject like interstellar turbulence is to use a synergistic approach, combining theoretical knowledge, numerical simulations, and observational data via statistical studies and data visualization tools.
For example, the turbulent power spectrum is often used to compare observations with both theoretical scaling laws and  numerical simulations. The Kolmogorov phenomenological description of turbulence has been indispensable for comparison of natural phenomena and has many applications in meteorology, engineering, etc.  

\begin{figure}
\includegraphics[width=8.5cm]{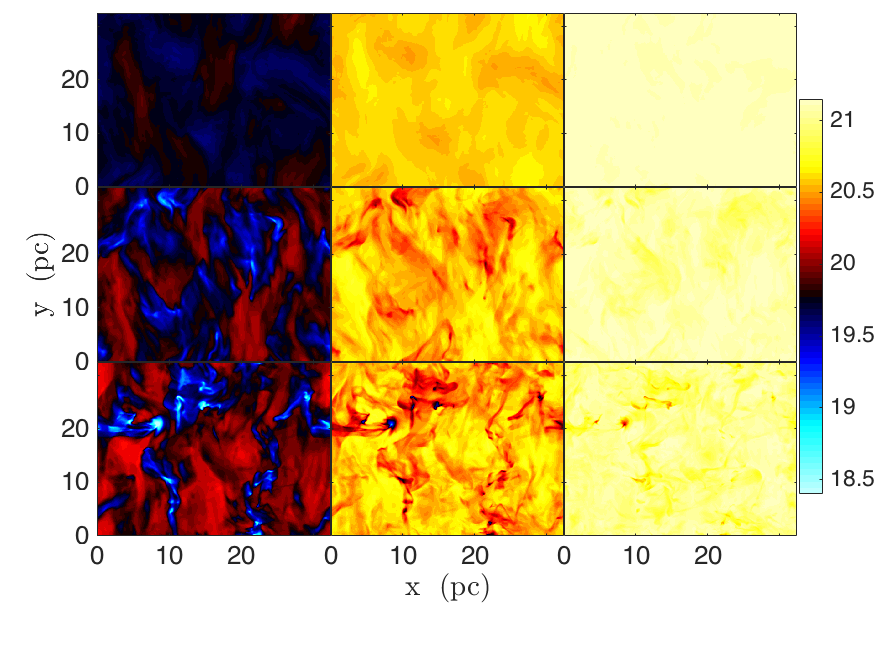}
\caption{
\label{fig:hi}
Reproduced from \cite{burkhart2020}: examples of synthetic HI column density maps produced from MHD turbulence simulations as part of the Catalog for Astrophysical Turbulence Simulations data release.  The color bar is $\log_{10}$($N_{\rm HI}$), and the 9~panels correspond to models with $\mathcal{M}_{\rm s}=0.5$, $2$, $4.5$ (from top row to bottom row, respectively), and interstellar radiation field $I_{\rm UV}=0.1$, $1$, $10$ (first to last column, respectively). }
\end{figure}

The importance of a synergistic comparison of theory, numerical simulations and observations can be understood best when analyzing the progress in a particular sub-field of the turbulence problem.  As an example, here we will highlight progress in understanding the nature of magnetic reconnection and diffusion of magnetic fields from star forming regions that has been made by careful comparison of theory, numerical simulations and observations. To start, consider the textbook concept of magnetic flux freezing introduced by \citet{Alfven1942}: plasma particles are entrained to the same magnetic field lines and bound to follow these field lines essentially indefinitely in the limit of zero plasma resistivity. This is known as the  ``frozen-in" condition and should be satisfied for quiescent astrophysical flows that have high conductivity.  However, nature seemingly violates this condition all the time. Solar flares are a well-known example of magnetic reconnection where the magnetic field topology changes and plasma is no longer bound to the magnetic field.  

Additionally, all MHD turbulence numerical simulations violate flux freezing due to finite resolution and numerical resistivity. 
Through a combination of numerical, theoretical and observational work, it is now clear that fluid turbulence plays the key role in solving the ``flux freezing paradox." In the model of turbulent magnetic reconnection proposed by \citet{LV99}, turbulence makes magnetic reconnection both fast and independent of plasma resistivity. This means that in the inertial range of the turbulent cascade,  the violation of the ``frozen-in" condition in high conductivity astrophysical flows is expected.  In practical terms, this theoretical insight means that, in terms of magnetic flux, numerical simulations reproduce physics correctly if the plasma is turbulent at the scales of interest.  This conclusion was confirmed in \citet{Eyin13}, where it was shown that the diffusion of the magnetic field in simulations is governed by the properties of turbulent motions and not by the plasma resistivity or numerical resistivity. Furthermore, observations of magnetic diffusion in molecular clouds suggest that magnetic reconnection is likely the dominate process of removing magnetic flux from collapsing regions on scales of 0.1 parsecs and greater \citep{Crutcher12a}. This provides justification for many numerical simulations of turbulent astrophysical environments, but suggests that a note of caution regarding attempts to analyze the results of MHD simulations at scales near the turbulent dissipation or driving.

\subsection{Open Source Tools for Studying Turbulence in Astrophysical Environments}

As discussed in detail in Section \ref{sec:stats}, a promising development  is the number of advances made over the last decade in new statistical diagnostics of turbulence. 
The recently developed statistical tools described above aim to capture the properties of the cascade, including the injection and dissipation mechanisms, the strength and properties of the magnetic field (encapsulated by the Alfv\'enic Mach number, $M_A$) and the compressibility of the medium (encapsulated by the sonic Mach number, $M_s$). 
 These statistics can be applied to observations such as column density maps, spectral line data (e.g., PPV cubes) or synchrotron polarization maps.

 However, until recently there was no common framework for the different turbulence statistics reviewed here.  
 \turbustat\ is a publicly available Python package that implements fourteen observational diagnostics of ISM turbulence described in this review and elsewhere in the literature \citep{koch2019AJ....158....1K}. \turbustat\ provides a common framework for running and comparing turbulence diagnostics, including comparisons between simulations and observations \citep[see][for some examples]{boyden2016ApJ...833..233B,boyden2018ApJ...860..157B,Koch2017,koch2019AJ....158....1K}.

Finally, we note that shared community resources can provide a way forward for different groups approaching the problem of astrophysical turbulence from a variety of perspectives. 
Shared simulation resources allow for increased collaboration and more cross-talk between observers and theorists or between researchers in traditionally different sub-fields working on turbulence.

A recent turbulence focused simulation database available to the astronomy community is the Catalog for Astrophysical Turbulence Simulations (CATS) found at www.mhdturbulence.com \citep{burkhart2020}. CATS includes 3D turbulence simulations from multiple codes/groups.  Synthetic observation data products can also be obtained, including turbulence-based polarization maps, PPV cubes and HI intensity maps. 
Perhaps the most notable turbulence simulation
database is the Johns Hopkins Turbulence Database (Li
et al. 2008). This database's primary audience is found in the 
incompressible turbulence community and not on supersonic MHD and self-gravitating turbulence such
as is found in the interstellar medium (ISM) of galaxies.
Another open simualtion database, the Starformat Simulations \footnote{ http://starformat.obspm.fr/}, includes
supersonic simulations and its simulations are primarily tailored for star forming GMC environments.

For studies of galactic magnetic fields, two exciting open-source endeavors are Galmag and IMAGINE. 
Galmag \citep{2019ascl.soft03005R} computes galactic magnetic fields based on mean field dynamo theory. Written in Python, Galmag allows efficient exploration of solutions to the mean field dynamo equation based on galaxy parameters specified by the user, such as the scale height profile, gas velocity dispersion, and the galaxy rotation curve. 

In a similar vein, the IMAGINE Consortium \citep{2019Galax...7...17H} aims to provide the community an enhanced and open-framework modeling of the magnetic field of the Milky Way. IMAGINE includes an open-source modular software pipeline that optimizes parameters in a user-defined galactic magnetic field model against various selected observational datasets.

 \subsection{The Rise of the Machines}
 We close our our discussion of turbulence tools with a look towards the revolution in machine learning techniques (MLT), which has already begun to dominate the statistical landscape and shows great promise for characterizing turbulence in observations.  The human capacity to visually distinguish features, including physically meaningful structures, is a long-coveted goal of 
computer and machine vision \citep{2015Natur.521..436L}.  The topic of this  subsection is vast enough to merit its own separate review.  CNNs  have been trained to differentiate simulations that are sub/super-Alfv\'{e}nic  \citep{Peek:2019:ApJL:}, feedback from stars \citep{2020ApJ...890...64X}, and  velocity fields from MHD simulations \citep{AsensioRamos:2017:A&A:}. However, here we will highlight a few  recent studies with the goal of introducing the reader to how MLT are being used in studies of interstellar turbulence. 

The rise of neural networks, deep learning, convolutional neural networks (CNNs), graph neural networks (GNNs), and generative adversarial networks (GANs) has completely revolutionized our ability to extract information from images and data cubes and even generate learned data. These techniques allow for hierarchical feature extraction and generalization and are now widely used for classification of astronomical images and data.  Indeed, almost 500 papers using neural networks in the astronomy and physics literature were  published in 2019.  While many of these articles are exploratory in nature, several apply neural networks to astronomical observations for the purpose of discovery. A more complete description of CNNs appropriate for an astronomical readership can be found in  \cite{Dieleman2015MNRAS.450.1441D}.

In particular, CNNs are a natural choice for classification problems and lend themselves well to being trained on MHD simulations.
CNNs are a subclass of neural networks that have a specific limited connectivity, originally designed to mimic visual neural structure in the human brain \citep{2014arXiv1412.6806S}. One of their first applications to astronomical images was in \cite{Dieleman2015MNRAS.450.1441D}, which used a CNN to reproduce visual classifications of galaxy morphology. 
In another example,  \citet{Peek:2019:ApJL:} trained a neural network to distinguish between two MHD turbulence simulations with different levels of magnetization.  They found that, even given a tiny slice of simulation data, a relatively simple CNN can distinguish between sub-Alfv\'enic (strong magnetic field) and super-Alfv\'enic (weak magnetic field) turbulence $>$98\% of the time, even if the power spectral amplitude information is removed from the data.

Another promising study by  \citet{2020MNRAS.tmp.2590T} shows that CNNs can be trained to determine  the spectral slope of turbulence in observations. These promising  preliminary studies are paving the way for the use of neural networks in uncovering properties of turbulence in real data, where observational biases and other issues need to be taken into account.

Comparing the results of statistical techniques of turbulence, simulations and analytic predictions with observations has thus far given the astronomy community a path forward when dealing with the turbulence problem. Machine learning will continue this progress forward into the near future.  In addition, it should be stressed that the 'glue' between these different areas will continue to be understanding the basic MHD turbulence scaling relations.  MHD scalings critically inform on how turbulence affects various astrophysical applications.
Communication between different sub-fields in astronomy and sharing of resources (e.g. the CATS, IMAGINE,  \turbustat\ projects) will ensure that the most important unsolved problem of classical physics will not slow down the impressive pace of astrophysical turbulence research. 

\acknowledgments{ \small
B.B. thanks the anonymous referee and Alex Lazarian for helpful comments on this review. B.B. is grateful to Naomi Klarreich for her helpful edits to the manuscript.    }

\bibliography{refs.bib}

\end{document}